\documentclass[MSNbibl,number,seceqn,citesort,dvips]{arxstspdf}
\usepackage{mathbh}
\usepackage{flushend}
\usepackage{stfloats}

\volume{27}
\issue{1}
\pubyear{2012}
\firstpage{61}
\lastpage{81}
\doi{10.1214/11-STS380}

\makeatletter
\newtheorem{theorem}{Theorem}[section]
\newtheorem{lemma}{Lemma}[section]
\newtheorem{corollary}{Corollary}[section]

\def\R{{\mathbb R}}
\newcommand{\cal}{\mathcal}
\newcommand{\one}{\mathbh{1}}
\newcommand{\nbOne}{\mathbh{1}}

\newcommand{\fracc}[2]{{#1}/{(#2)}}
\newcommand{\fraca}[2]{{#1}/{#2}}
\renewcommand{\*}{\cdot\allowbreak}
\makeatother

\begin{document}
\begin{frontmatter}
\vspace*{6pt}
\title{On Improved Loss Estimation for~Shrinkage Estimators}
\runtitle{Loss Estimation}

\begin{aug}
\author[a]{\fnms{Dominique} \snm{Fourdrinier}\ead[label=e1]{dominique.fourdrinier@univ-rouen.fr}}
\and
\author[b]{\fnms{Martin T.} \snm{Wells}\corref{}\ead[label=e2]{mtw1@cornell.edu}}
\runauthor{D. Fourdrinier and M. T. Wells}

\affiliation{Universit\'e de Rouen and Cornell University, and Cornell
University}

\address[a]{Dominique Fourdrinier is Professor, Universit\'e de Rouen,
LITIS EA 4108,
Avenue de l'Universit\'e, BP 12, 76801 Saint-\'Etienne-du-Rouvray,
France,
and Visiting Professor, Department of Statistical
Science, Cornell University,
Comstock Hall, Ithaca, New York 14853, USA \printead{e1}.}
\address[b]{Martin T. Wells is Professor, Department
of Statistical
Science, Cornell University,
Comstock Hall, Ithaca, New York 14853, USA \printead{e2}.}

\end{aug}

%
\begin{abstract}
Let $X$ be a random vector with distribution $P_\theta$
where $\theta$ is an unknown parameter. When estimating $\theta$ by
some estimator $\varphi(X)$ under a loss function
$L(\theta,\varphi)$, classical decision theory advocates that such a
decision rule should be used if it has suitable properties with
respect to the frequentist risk $R(\theta,\varphi)$. However, after
having observed $X = x$, instances arise in practice in which
$\varphi$ is to be accompanied by an assessment of its loss,
$L(\theta,\varphi(x))$, which is unobservable since $\theta$ is
unknown. A common approach to this assessment is to consider
estimation of $L(\theta,\varphi(x))$ by an estimator $\delta$,
called a loss estimator. We present an expository development of
loss estimation with substantial emphasis on the setting where the
distributional context is normal and its extension to the case where
the underlying distribution is spherically symmetric. Our overview
covers improved loss estimators for least squares but primarily focuses
on shrinkage estimators. Bayes estimation is also considered and
comparisons are made with unbiased estimation.
\end{abstract}

%
\begin{keyword}
\kwd{Conditional inference}
\kwd{linear model}
\kwd{loss estimation}
\kwd{quadratic loss}
\kwd{risk function}
\kwd{robustness}
\kwd{shrinkage estimation}
\kwd{spherical symmetry}
\kwd{SURE}
\kwd{unbiased estimator of loss}
\kwd{uniform distribution on a sphere}.
\end{keyword}

\end{frontmatter}

\section{Introduction} \label{Section1}\vspace*{-2pt}
Suppose $X$ is an observable from a distribution~$P_\theta$
parameterized by an unknown
parameter $\theta$. In classical decision theory, it is usual, after
selecting an estimation
procedure $\varphi(X)$ of $\theta$, to evaluate it through a loss criterion,
$L(\theta,\varphi(X)$, which represents the cost incurred by the
estimate $\varphi(X)$ when
the unknown parameter equals~$\theta$. In the long run, as it depends
on the particular value
of $X$, this loss cannot be appropriate to assess the performance of
the estimator $\varphi$.
Indeed, to be valid (in the frequentist sense), a global evaluation of
such a statistical procedure
should be based on averages over all the possible observations.
Consequently, it is common to report the risk
$R(\theta,\varphi) = E_\theta[L(\theta,\varphi(X)]$ as a measure
of the efficiency of $\varphi$
($E_\theta$ denotes expectation with respect to $P_\theta$). Thus we
have at our disposal a
long-run performance of~$\varphi(X)$ for each value of~$\theta$.
However, although this notion
of risk can effectively be used in comparing $\varphi(X)$ with other
estimators, it is
inaccessible since $\theta$ is unknown. The usual frequentist risk
assessment is the maximum risk
$\overline{R}_\varphi= \sup_\theta R(\theta,\varphi)$.

By construction, this least favorable report of the estimation
procedure is non-data-dependent [as we
were guided by a global notion of accuracy of $\varphi(X)$]. However,
there exist situations
where the fact that the observation $X$ has such or such value may
influence the judgment on
a statistical procedure. A particularly edifying example is given by
the following simple
confidence interval estimation (which can be viewed as a loss
estimation problem). Assume that the observable is a couple $(X_1,X_2)$
of independent copies of a random
variable $X$ satisfying, for $\theta\in\R$,
\[
P[X = \theta-1] = P[X = \theta+1] = \tfrac{1}{2}   .
\]
Then it is clear that the confidence interval for $\theta$ defined
by
\[
I(X_1,X_2) =
 \biggl\{\theta\in\R\Bigm| \biggl|\frac{X_1+X_2}{2} - \theta \biggr|
< \frac{1}{2} \biggr\}
\]
satisfies
\[
\one_{[\theta\in I(X_1,X_2)]} = \cases{
1, &  if    $X_1 \not= X_2$, \cr
0, &  if    $X_1 = X_2$,
}\]
so that it suffices to observe $(X_1,X_2)$ in order to know exactly
whether $I(X_1,X_2)$ contains $\theta$ or not.

The previous (ad hoc) example indicates that da\-ta-dependent reports
are relevant. When $X = x$ the loss, $L(\theta,\varphi(x))$, itself
could serve as a perfect measure of the accuracy of $\varphi$ if it
were available (which it is not since $\theta$ is unknown). It is
natural to estima\-te~$L(\theta,\varphi(x))$ by a data-dependent
estimator $\delta(X)$,\break a~new estimator called a loss estimator. Such
an estimator can serve as a data-dependent assessment (instead of
$\overline{R}_\varphi$). This is a conditional approach in~the sense that the accuracy
assessment is made on a~da\-ta-dependent quantity,
the loss, instead of the risk.

To evaluate the extent to which $\delta(X)$ successfully estimates
$L(\theta,\varphi(X))$,
another loss is required and it has become standard, for simplicity, to
use the squared error
%
\begin{equation} \label{squaredloss}
 \qquad L^*(\theta,\varphi(X),\delta(X)) = \bigl(\delta(X) - L(\theta,\varphi
(X))\bigr)^2
.
\end{equation}
Insofar as we are thinking in terms of long-run frequencies, we
adopt
a frequentist approach to evaluating the performance of $L^*$ by averaging
over the sampling
distribution of $X$ given $\theta$, that is, by using a new notion of risk
%
\begin{eqnarray} \label{newrisk}
{\cal R(\theta,\varphi,\delta)} &=& E_\theta[L^*(\theta,\varphi
(X),\delta(X))]\nonumber
\\[-8pt]
\\[-8pt]
&=& E_\theta\bigl[\bigl(\delta(X) - L(\theta,\varphi(X))\bigr)^2\bigr]   .
\nonumber
\end{eqnarray}
As $\overline{R}_\varphi$ reports on the worst possible situation
(the maximum risk), we may expect
that a competitive data-dependent report $\delta(X)$ should improve on
$\overline{R}_\varphi$
under the risk (\ref{newrisk}), that is, for all $\theta$, $\delta
(X)$ satisfies
%
\begin{equation} \label{improve3}
{\cal R(\theta,\varphi,\delta)} \leq{\cal R}(\theta,\varphi
,\overline{R}_\varphi)   .
\end{equation}
More generally, a reference loss estimator $\delta_0$ will be
dominated by a competitive
estimator $\delta$ if, for all $\theta$,
%
\begin{equation} \label{improve4}
{\cal R(\theta,\varphi,\delta)} \leq{\cal R}(\theta,\varphi
,\delta_0)   ,
\end{equation}
with strict inequality for some $\theta$.

Unlike the usual estimation setting where the quantity of interest
is a function of the parameter $\theta$, loss estimation involves a
function of both~$\theta$ and~$X$ (the data). This feature may make
the statistical analysis more difficult but it is clear
that the usual notions of minimaxity, admissibility, etc., and their
methods of proof can be
directly adapted to that situation. Also, although frequentist
interpretability was evoked
above, in case we would be interested in a Bayesian approach, it is
easily seen that this
approach would consist of the usual Bayes estimator $\varphi_B$ of
$\theta$ and the posterior
loss $\delta_B(X) = E[L(\theta,\varphi_B)|X]$.

The problem of estimating a loss function has been considered by
Sandved \cite{Sandved1968}
who developed a~notion of unbiased estimator of $L(\theta,\varphi
(X))$ in various settings.
However, the underlying conditional approach traces back to Lehmann and
Sheff\'e
\cite{LehmannSheffe1950} who estimated the power of a statistical
test. Kiefer, in a series
of papers \cite{Kiefer1975,Kiefer1976,Kiefer1977},
developed conditional
and estimated confidence theories. A subjective Bayesian
approach was compared by Berger \cite{Berger1985b,Berger1985c,Berger1985a}
with the frequentist paradigm. Jonhstone \cite{Johnstone1988}
considered (in)ad\-missibility of
unbiased estimators of loss for the ma\-ximum likelihood estimator
$\varphi_0(X) = X$ and for\break
the~James--Stein estimator $\varphi^{\mathrm{JS}}(X) =  (1-(p-2)
/\break
\|X\|^2 ) X$ of a~$p$-variate normal mean $\theta$.
For $\varphi_0(X) = X$, the unbiased estimator of the quadratic
loss\break
$L(\theta,\varphi_0(X)) = \|\varphi_0(X)-\theta\|^2$, that is, the
loss estimator
$\delta_0$ which satisfies, for all $\theta$,
%
\begin{equation} \label{lambda0}
E_\theta[\delta_0] = E_\theta[L(\theta,\varphi_0(X))] = R(\theta
,\varphi_0)   ,
\end{equation}
is $\delta_0 = \overline{R}_\varphi= p$. Johnstone proved that (\ref
{improve3}) is satisfied
with the competitive estimator $\delta(X) = p   -2(p-4)/\|X\|^2$ when
$p \geq5$, the
risk difference between $\delta_0$ and $\delta$ being expressed as
$-4(p-4)^2 E_\theta[1/\|X\|^4]$. For the James--Stein estimator
$\varphi^{\mathrm{JS}}$, the unbiased estimator
of loss is itself
data-dependent and equal to
$\delta_0^{\mathrm{JS}}(X) = p   -(p-2)^2/\|X\|^2$. Jonhstone showed that
improvement on
$\delta_0^{\mathrm{JS}}$ can be obtained with
$\delta^{\mathrm{JS}}(X) = p   -(p-2)^2/\|X\|^2 + 2p/\|X\|^2$
when $p \geq5$, with strict inequality in (\ref{improve4}) for all
$\theta$ since the
difference in risk between $\delta^{\mathrm{JS}}$ and $\delta_0^{\mathrm{JS}}$ equals
$-4p^2 E_\theta[1/\|X\|^2]$.

In Section \ref{Section2}, we develop the quadratic loss estimation
problem for a $p$-normal
mean. After a review of the basic ideas, a new class of loss estimators
is constructed in
Section \ref{subsection2.1}. In Section \ref{subsection2.2}, we
turn our focus on some
interesting and surprising behavior of Bayesian assessments; this
paradoxical result is
illustrated in a general inadmissibility theorem. Section~\ref
{Section3} is devoted to the
case where the variance is unknown. Extensions to the spherical case
are given in Section~\ref{Section4}. In  Section \ref{subsection4.1}, we consider the
general case of a
spherically symmetric distribution around a fixed vector $\theta\in\R
^p$ and in
 Section \ref{subsection4.2} these ideas are then generalized to the
case where a residual
vector is available. We conclude by mentioning a~number of applied and
theoretical developments
of loss estimation not covered in this overview. The \hyperref[appendix]{Appendix} gives
some necessary background material
and technical results.


\section{Estimating the Quadratic Loss of a $p$-Normal Mean with Known
Variance}
\label{Section2}

\subsection{Dominating Unbiased Estimators of Loss} \label{subsection2.1}
Let $X$ be a $p$-variate normally distributed ${\cal N}(\theta,I_p)$
random vector with unknown mean $\theta$ and identity covariance
matrix $I_p$. To estimate $\theta$, the
observable~$X$ is itself a reference estimator (it is the maximum
likelihood estimator (m.l.e.)
and it is an unbiased estimator of $\theta$) so that it is convenient
to write any
estimator of $\theta$ through $X$ as $\varphi(X) = X+g(X)$, for a
certain function $g$ from
$\R^p$ into $\R^p$. Under squared error loss $\|\varphi(X)-\theta
\|^2$, the (quadratic)
risk of $\varphi$ is defined by
%
\begin{equation} \label{risk}
R(\theta,\varphi) = E_\theta[\|\varphi(X)-\theta\|^2],
\end{equation}
where $E_\theta$ denotes the expectation with respect to ${\cal
N}(\theta,I_p)$.

Clearly, the risk of the m.l.e. $X$ equals $p$ and in general
$\varphi(X)$ will be a reasonable estimator
only if its risk is finite. It is easy to see (Lemma~\ref{LemmaA2}  in Appendix~\ref{A.1}) through Schwarz's inequality that this is the
case as soon as
%
\begin{equation} \label{finitness1}
E_\theta[\|g(X)\|^2] < \infty  ,
\end{equation}
which we will assume in the following (it can be also seen that this
condition is in fact
necessary to guarantee the risk finiteness).

To improve on the m.l.e. $X$ when $p \geq3$ [i.e.,\break to have $R(\theta
,\varphi) \leq p$], Stein
\cite{Stein1981} exhibited (under certain differentiability conditions
that we recall
below) an unbiased estimator of the risk of $\varphi(X)$, that is, a~function $\delta_0(X)$
(depending only on $X$ and not on~$\theta$) for which
%
\begin{equation} \label{unbrisk}
R(\theta,\varphi) = E_\theta[\delta_0(X)]   .
\end{equation}
This suggests a natural estimator of the loss
$\|\varphi(X)-\theta\|^2$ since
(\ref{unbrisk}) implies that
%
\begin{equation} \label{unbloss}
E_\theta[\|\varphi(X)-\theta\|^2] = E_\theta[\delta_0(X)]
\end{equation}
and hence is an unbiased estimator of the loss.\break Stein~\cite{Stein1981} proved more precisely
that $\delta_0(X) = p +\break 2 \*  \operatorname{div}g(X) + \|g(X)\|^2$
[where $\operatorname{div}g(X)$ stands
for the divergence of $g(X)$, i.e., $\operatorname{div}g(X) = \sum
_{i=1}^p \partial_ig_i(X)$]. One
can see that $\delta_0$ may change sign so that, as an estimator of
loss (which is nonnegative), it cannot be completely satisfactory, and
hence, is likely to be improved upon.

Any competitive loss estimator $\delta(X)$ can be written as
$\delta(X) = \delta_0(X) - \gamma(X)$ for a certain func-\break tion~$\gamma
(X)$ which can be
interpreted as a correction to~$\delta_0(X)$. Note that, for the m.l.e.
[i.e., if $g(X)=0$], we may expect that an improvement on \mbox{$\delta_0(X)
= p$} would be obtained
with a nonnegative function $\gamma(X)$ satisfying the requirement
expressed by condi-\break tion~(\ref{improve3}). Note also that, similarly to the finiteness risk condition
(\ref{finitness1}), we will require that
%
\begin{equation} \label{finitness2}
E_\theta[\gamma^2(X)] < \infty
\end{equation}
to assure that the risk of $\delta(X)$ is finite (see Appendix~\ref{A.1}).

Using straightforward algebra, the risk difference
${\cal D}(\theta,\varphi,\delta) =
{\cal R}(\theta,\varphi,\delta) - {\cal R}(\theta,\varphi,\delta_0)$
simplifies to
%
\begin{eqnarray} \label{diffrisk}
{\cal D}(\theta,\varphi,\delta) &=& E_\theta[\gamma^2(X) - 2
\gamma(X)   \delta_0(X)]\nonumber
\\[-8pt]
\\[-8pt]
&&{}+ 2   E_\theta[\gamma(X)   \|\varphi(X) - \theta\|^2]   .
\nonumber
\end{eqnarray}
Conditions for which ${\cal D}(\theta,\varphi,\delta) \leq0$ will
be formulated after
finding an unbiased estimate of the term
$\gamma(X)   \|\varphi(X) - \theta\|^2$ in the last expectation. We
briefly review the
flow of ideas of those techniques.

For a function $g$ from $\R^p$ into $\R^p$, the Stein's identity (see
Stein \cite{Stein1981})
states that
%
\begin{equation} \label{SteinIdentity}
E_\theta[(X-\theta)^tg(X)] = E_\theta[\operatorname{div}g(X)]
\end{equation}
provided that these expectations exist. Here Stein specified that $g$
was almost
differentiable. Weak differentiability is needed to integrate shrinkage
functions $g(X)$,
intervening in the James--Stein estimators, of the form $g(X) = -a   X
  / \|X\|^2$ which are
not differentiable in the usual sense [such a $g(X)$ explodes at zero].
This notion is equivalent
(and it is of more common use in analysis) to the statement that $g$
belongs to the Sobolev
space $W_{\mathrm{loc}}^{1,1}(\R^p)$ of weakly differentiable functions. That
equivalence was
noticed by Johnstone \cite{Johnstone1988}.

Recall that a locally integrable function $\gamma$ from~$\R^p$ into~$\R$ is said\vadjust{\goodbreak} to be weakly
differentiable if there exist~$p$ functions $h_1, \ldots,h_p$ locally
integrable on $\R^p$
such that, for any $i=1, \ldots,p$,
%
\begin{equation} \label{weakdiff}
 \quad \int_{\R^p} \gamma(x)   \frac{\partial\varphi}{\partial x_i}(x)
\, dx
=
- \int_{\R^p} h_i(x)   \varphi(x) \, dx
\end{equation}
for any infinitely differentiable function $\varphi$ on $\R^p$ with
compact support. The
functions $h_i$ are the $i$th partial weak derivatives of $\gamma$.
Their common notation is
$\partial\gamma/ \partial x_i$ and the vector
$\nabla\gamma= (\partial\gamma/ \partial x_1, \ldots,\partial
\gamma/\break\partial x_p)^t$
is referred to as the weak gradient of $\gamma$.

Note that (\ref{weakdiff}) usually holds when $\gamma$ is
continuously differentiable, that is,
when $h_i = \partial\gamma/ \partial x_i$, the standard partial
derivative, is continuous.
Thus,\break via~(\ref{weakdiff}), the extension to weak differentiability
consists in a propriety of
integration by parts with vanishing bracketed term. Naturally a function
$g = (g_1,\break\ldots,g_p)$ from $\R^p$ into $\R^p$ is said to be weakly
differentiable if each of
its components $g_j$ is weakly differentiable. In that case, the function
$\operatorname{div}g = \sum_{i=1}^p \partial g_i /\break\partial x_i$ is
referred to as the weak
divergence of $g$; this is the operator intervening in the Stein's identity
(\ref{SteinIdentity}).

When dealing with an unbiased estimator of\break a~quantity of the form
$\|X-\theta\|^2   \gamma(X)$,
where $\gamma$ is a~function from $\R^p$ into $\R$, writing
%
\begin{equation} \label{norme2}
\|X-\theta\|^2   \gamma(X) = (X-\theta)^t   (X-\theta)   \gamma(X)
\end{equation}
naturally leads to an iteration of Stein's identi-\break ty~(\ref
{SteinIdentity}) and involves twice
weak differentiability of~$\gamma$. This is of course defined through
the weak
differentiability of all the weak partial derivatives $\partial\gamma
/ \partial x_i$;
these second weak partial derivatives are denoted by
$\partial^2 \gamma/ \partial x_j\, \partial x_i$. Thus $\gamma$
belongs to the Sobolev
spa-\break ce~$W_{\mathrm{loc}}^{2,1}(\R^p)$ and
$\Delta\gamma= \sum_{i=1}^p \partial^2 \,\gamma/ \partial x_i^2$
is referred to as the weak Laplacian of $\gamma$.

By (\ref{norme2}) and (\ref{SteinIdentity}), we have
%
\begin{eqnarray} \label{unbiasednorme2}
 &&E_\theta[\|X-\theta\|^2   \gamma(X)]\nonumber\\  && \quad = E_\theta\bigl[\operatorname
{div}\bigl((X-\theta)^t   \gamma(X)\bigr)\bigr]
\\
&& \quad    = E_\theta[p   \gamma(X) + (X-\theta)^t   \nabla\gamma(X)]
\nonumber
\end{eqnarray}
by the product rule for the divergence operator. Then, applying again
(\ref{SteinIdentity}) to the last term in (\ref{unbiasednorme2})
gives
%
\begin{eqnarray} \label{Laplunbiased}
E_\theta[(X-\theta)^t   \nabla\gamma(X)] &=& E_\theta[\operatorname
{div}(\nabla\gamma(X)]\nonumber
\\[-8pt]
\\[-8pt]
&=& E_\theta[\Delta\gamma(X)]
\nonumber
\end{eqnarray}
by definition of the Laplacian operator. Finally, gathering
(\ref{unbiasednorme2}) and
(\ref{Laplunbiased}), we have that
%
\begin{eqnarray} \label{unbiasednorme22}
&&E_\theta[\|X-\theta\|^2   \gamma(X)] \nonumber
\\[-8pt]
\\[-8pt]
&& \quad = E_\theta[p   \gamma(X) +
\Delta\gamma(X)]   .
\nonumber
\end{eqnarray}

We are now in a position to provide an unbiased estimator of the
difference\vadjust{\goodbreak} in risk
${\cal D}(\theta,\varphi,\delta)$ in (\ref{diffrisk}). Its
nonpositivity will be a
sufficient condition for ${\cal D}(\theta,\varphi,\delta) \leq0$
and hence for $\delta$
to improve on $\delta_0$. Indeed we have
\begin{eqnarray*}
&&\|\varphi(X) - \theta\|^2\\
 && \quad = \|X + g(X) - \theta\|^2 \\
&& \quad = \|g(X)\|^2 + 2   (X-\theta)^t   g(X) + \|X - \theta\|^2
\end{eqnarray*}
so that, according to (\ref{SteinIdentity}) and (\ref{unbiasednorme22}),
\begin{eqnarray*}
&&E_\theta[\|\varphi(X)-\theta\|^2   \gamma(X)]\\
&& \quad  =
E_\theta[\gamma(X)   \|g(X)\|^2 + 2   \operatorname{div}(\gamma
(X)   g(X))\\
&&\hspace*{105pt}+ p   \gamma(X) + \Delta\gamma(X)]   .
\end{eqnarray*}
Therefore, as $ \operatorname{div}(\gamma(X)   g(X)) = \gamma(X)
\operatorname{div}g(X) +\break  \nabla\gamma(X)^t   g(X)$
and as $\delta_0(X) = p + 2   \operatorname{div}g(X) + \break \|g(X)\|^2$,
the risk difference
${\cal D}(\theta,\varphi,\delta)$ in (\ref{diffrisk}) reduces to
\[
{\cal D}(\theta,\varphi,\delta) =
E_\theta[\gamma^2(X) + 4   \nabla\gamma(X)^t   g(X) + 2   \Delta
\gamma(X)],
\]
so that a sufficient condition for ${\cal D}(\theta,\varphi,\delta)$
to be nonpositive is
%
\begin{equation} \label{domincond}
\gamma^2(x) + 4   \nabla\gamma(x)^t   g(x) + 2   \Delta\gamma
(x) \leq0
\end{equation}
for any $x \in\R^p$.

The question now arises of determining a ``best'' correction $\gamma$
satisfying
(\ref{domincond}). The following theorem
provides a way to associate to the function $g$ a suitable correction
$\gamma$ which satisfies
(\ref{domincond}) in the case where $g(x)$ is of the form $g(x) =
\nabla m(x) / m(x)$ for a
certain nonnegative function $m$. This is the case when $\varphi$ is a
Bayes estimator of
$\theta$ related to a prior~$\pi$, the function $m$ being the
corresponding marginal (see
Brown \cite{Brown1971}). Bock \cite{Bock1988} showed that, through
the choice of $m$, such
estimators constitute a wide class of estimators of $\theta$ (which
are called pseudo-Bayes
estimators when the function $m$ does not correspond to a true prior
$\pi$).

\begin{theorem}\label{th2.1}
Let $m$ be a nonnegative function which is also superharmonic
(respectively subharmonic) on
$\R^p$ such that $\nabla m / m \in W_{\mathrm{loc}}^{1,1}(\R^p)$. Let $\xi$
be a real-valued function,
strictly positive and strictly subharmonic (respectively superharmonic)
on $\R^p$\break such that
%
\begin{equation} \label{finitness}
E_\theta \biggl[ \biggl(\frac{\Delta\xi(X)}{\xi(X)} \biggr)^2
\biggr] < \infty  .
\end{equation}
Assume also that there exists a constant $K>0$ such that, for any $x
\in\R^p$,
%
\begin{equation} \label{marginalup}
m(x) > K \frac{\xi^2(x)}{|\Delta\xi(x)|}
\end{equation}
and let $K_0 = \inf_{x \in\R^p} m(x)
\frac{|\Delta\xi(x)|}{\xi^2(x)}$.\vadjust{\goodbreak}

Then the unbiased loss estimator $\delta_0$ of the estimator $\varphi
$ of $\theta$ defined by
$\varphi(X) = X + \nabla m(X) / m(X)$
is dominated by the estimator $\delta= \delta_0 - \gamma$, where the
correction term
$\gamma$ is given, for any $x \in\R^p$ such that $m(x) \not= 0$, by
%
\begin{equation} \label{correction}
\gamma(x) = - \alpha  \operatorname{sgn}(\Delta\xi(x))   \frac
{\xi(x)}{m(x)}   ,
\end{equation}
as soon as $0 < \alpha<2   K_0$.
\end{theorem}

\begin{pf} The domination condition will be shown by
proving that the risk difference is less than zero.
We only consider the case where $m$ is superharmonic and $\xi$ is
strictly subharmonic, the
case where $m$ is subharmonic and $\xi$ is strictly superharmonic
being similar.

First note that the finiteness risk condition (\ref{finitness2}) is
guaranteed by the condition in
(\ref{finitness}) and the fact that (\ref{marginalup}) implies that,
for any $x \in\R^p$,
\[
\gamma^2(x) =
\alpha^2   \frac{\xi^2(x)}{m^2(x)} \leq
\frac{\alpha^2}{K_0^2}   \biggl (\frac{\Delta\xi(x)}{\xi
(x)} \biggr)^2   .
\]

Further note that, for a shrinkage function $g$ of the form $g(x) =
\nabla m(x) / m(x)$, the left-hand side of~(\ref{domincond}) can be
expressed as
%
\begin{eqnarray} \label{operator}
{\cal R}\gamma(x) &=& \gamma^2(x)\nonumber\hspace*{-35pt}
\\[-8pt]
\\[-8pt]
&&+ 2\biggl \{2\frac{\Delta(m(x)\gamma(x))}{m(x)} - \gamma(x) \frac{\Delta m(x)}{m(x)}
\biggr\}\hspace*{-35pt}
\nonumber
\end{eqnarray}
and hence, for $\gamma$ in (\ref{correction}), as
%
\begin{eqnarray} \label{operator2}
{\cal R}\gamma(x) &=& \alpha^2   \frac{\xi^2(x)}{m^2(x)}\nonumber\hspace*{-30pt}
\\[-8pt]
\\[-8pt]
&&+ 2\alpha\biggl\{- \frac{\Delta\xi(x)}{m(x)} + \frac{\xi(x)   \Delta m(x)}{m^2(x)} \biggr\} .\hspace*{-30pt}
\nonumber
\end{eqnarray}
Now, since $m$ is superharmonic and $\xi$ is positive, it follows from
(\ref{operator2}) that
\begin{eqnarray*}
{\cal R}\gamma(x) \leq
\frac{\alpha}{m(x)}  \biggl\{\frac{\alpha  \xi^2(x)}{m(x)} - 2
\Delta\xi(x) \biggr\}
\end{eqnarray*}
and hence, by subharmonicity of $\xi$, the inequality in~(\ref{marginalup})
and the definition of $K_0$,
that
%
\begin{equation} \label{operator3}
{\cal R}\gamma(x) <
\frac{\alpha}{m(x)} \{\alpha- 2   K_0\}   \frac{\xi^2(x)}{m(x)}
  .
\end{equation}
Finally, since $0 < \alpha<2   K_0$, the inequality in (\ref{operator3})
gives ${\cal R}\gamma(x) <0$, which is the desired result.
\end{pf}

As an example, consider $m(x) = 1/\|x\|^{p-2}$, that is, the
fundamental harmonic function which
is superharmonic on the entire space $\R^p$ (see Du Plessis~\cite
{Duplessis1970}).
Then we have $\nabla m(x) / m(x) = - (p-2) / \|x\|^2$\break and~$\varphi(X)$
is the James--Stein
estimator whose unbiased estimator of loss is $\delta_0(X) = p -
(p-2)^2 / \|X\|^2$.
First note that $\nabla m / m \in W_{\mathrm{loc}}^{1,1}(\R^p)$ for $p \geq3$.
Now choosing, for any
$x \not= 0$, the function $\xi(x) = 1 / \|x\|^p$ gives rise to
$\Delta\xi(x) = 2   p / \|x\|^{p+2} > 0$ and hence to
\[
\frac{\xi^2(x)}{|\Delta\xi(x)|} = \frac{1}{2   p} \frac
{1}{\|x\|^{p-2}}   ,
\]
which means that condition (\ref{marginalup}) is satisfied with $K < 2
  p$. Also we have
\[
 \biggl(\frac{\Delta\xi(x)}{\xi(x)} \biggr)^2 = \frac{4   p^2}{\|x\|^4}
\]
which implies that the condition in (\ref{finitness}) is satisfied for
$p \geq5$. Now it is clear
that the constant~$K_0$ is equal to $2   p$ and that the correction
term $\gamma$ in~(\ref{correction}) equals, for any $x \not= 0$, $\gamma(x) = -
\alpha/ \|x\|^2$. Finally,
Theorem \ref{th2.1} guarantees that an improved loss estimator over
the unbiased estimator of
loss $\delta_0(X)$ is
$\delta(X) = \delta_0(X) + \alpha/ \|x\|^2$ for $0 < \alpha< 4
p$, which is Johnstone's
result \cite{Johnstone1988} for the James--Stein estimator.

Similarly Johnstone's result for $\varphi(X) = X$ can be constructed
with $m(x) = 1$ (which is
both subharmonic and superharmonic) and with the choice of the
superharmonic function
$\xi(x) = 1/\|x\|^2$, for which $K_0 = 2   (p-4)$, so that $\delta
(x) = p - \alpha/\|x\|^2$
dominates~$p$ for $0 < \alpha< 4   (p-4)$.

We have shown that the unbiased estimator of loss can be dominated.
Often one may wish to add a~frequentist-validity constraint to a~loss estimation problem.
Specifically in our problem,
the frequentist-validity constraint for some estimator $\delta$ would be
$E_{\theta} [\delta(X)] \geq E_{\theta} [\delta_{0}(X)]$ for all
$\theta$.
Kiefer \cite{Kiefer1977} suggested that conditional
and estimated confidence assessments should be conservatively
biased, that is, the average reported loss should be greater than or
equal to
the average actual loss. Under such a frequentist-validity
condition Lu and Berger \cite{LuBerger1989} gave improved loss
estimators for several of the most important Stein-type estimators.
One of their estimators is a~ge\-neralized Bayes estimator,
suggesting that Bayesians and frequentists can potentially agree
on a conditional assessment of loss.

A possible problem with the improved estimator defined in (\ref
{correction}) is that
it may be negative, which is undesirable since we are estimating a
nonnegative quantity. A simple remedy to this problem is to use
a positive-part estimator. If we define the positive-part as
$\delta^+ = \max\{\delta, 0\}$, the loss difference between~$\delta^+$
and $\delta$ is $(\delta- L(\theta,\varphi))^2 -
(\delta^+
-L(\theta,\varphi))^2 = (\delta^2 -2\delta
L(\theta,\varphi))\nbOne_{\delta\le0}$, hence it is always
nonnegative. Therefore the risk difference is positive, which
implies that $\delta^+$ domimates~$\delta$. It would be of
interest to find an estimator that dominates $\delta^+$.

In the context of variance estimation, despite warnings on its
inappropriate behavior (Stein \cite{Stein1964},\break Brown~\cite{Brown1968}) the decision-theoretic approach to the normal
variance estimation is typically based on the standardized quadratic
loss function, where overestimation of the variance is much more
severely penalized than underestimation, thus leading to presumably
too small estimates. Similarly in loss estimation under quadratic
loss, the overestimation of the loss is also much more severely
penalized than underestimation. A possible alternative to quadratic
loss would be a Stein-type loss. Suppose $\varphi(X)$ is an
estimator of $\theta$ under $\| \theta-
\varphi(X)\| ^{2}$ and let $\delta(X)$ be an estimator of
$\| \theta- \varphi(X)\| ^{2}$ for $\delta(X) > 0$.
Then we can define the Stein-type loss for evaluating $\delta(X)$ as
%
\begin{eqnarray} \label{Steinloss}
L(\theta, \varphi(X), \delta(X)) &=& \frac{\| \theta-
\varphi(X)\| ^{2}}{\delta(X)} \nonumber
\\[-8pt]
\\[-8pt]&&{}- \log\frac{\| \theta-
\varphi(X)\| ^{2}}{\delta(X)} - 1.
\nonumber
\end{eqnarray}
The analysis
of the loss estimates under the Stein-type loss is more challenging but
can be carried out using the integration-by-parts tools developed in
this section.

\subsection{Dominating the Posterior Risk} \label{subsection2.2}

In the previous sections, we have seen that the unbiased estimator of
loss should be often
dismissed since it can be dominated. When a (generalized) Bayes
estimator of $\theta$ is
available, incorporating the same prior information for estimating the
loss of this Bayesian
estimator is coherent, and we may expect that the corresponding Bayes
estimator is a good
candidate to improve on the unbiased estimator of loss. However,
somewhat surprisingly,
Fourdrinier and Strawderman \cite{FourdrinierStrawderman2003} found that,
in the normal setting considered in  Section \ref{subsection2.1},
the unbiased
estimator often dominates the corresponding generalized Bayes estimator
of loss for priors
which give minimax estimators in the original point estimation problem.
They also gave a
general inadmissibility result for a generalized Bayes estimator of
loss. While much of their
focus is on pseudo-Bayes estimators, in this section, we essentially
present their results on
generalized Bayes estimators.

For a given generalized prior $\pi$, we denote the generalized
marginal by $m$ and the
generalized Bayes estimator of $\theta$ by
%
\begin{eqnarray} \label{equa3}
\varphi_m (X) = X + \frac{\nabla m (X)}{m (X)}   .
\end{eqnarray}
Then (see Stein \cite{Stein1981}) the unbiased estimator of risk of
$\varphi_m(X)$ is
%
\begin{eqnarray} \label{equa5}
\delta_0 (X) = p + 2 \frac{\Delta m (x)}{m (X)} - \frac{\| \nabla m
(X) \|^2}{m^2 (X)}
\end{eqnarray}
while the posterior risk of $\varphi_m(X)$ is
%
\begin{eqnarray}\label{equa6}
 \quad \delta_m (X) = p + \frac{\Delta m (X)}{m (X)} - \frac{\| \nabla m
(X) \|^2}{m^2 (X)}   .
\end{eqnarray}

Domination of $\delta_0 (X)$ over $\delta_m (X)$ is obtained\break thanks
to the fact that their
risk admits
$(\Delta m(X) /\break m(X))^2 - 2   \Delta^{(2)} m(X) / m(X)$
as an unbiased estimator of their risk difference, that is,
%
\begin{eqnarray} \label{equa11}
&&{\cal R} (\theta, \varphi_m, \delta_0) - {\cal R} (\theta, \varphi
_m, \delta_m)\nonumber
\\[-8pt]
\\[-8pt]
&& \quad =
E_\theta\biggl [
\biggl (\frac{\Delta m (X)}{m(X)} \biggr)^2 - 2   \frac{\Delta^{(2)}
m (X)}{m (X)}
 \biggr],
\nonumber
\end{eqnarray}
where $\Delta^{(2)} m = \Delta(\Delta m)$ is the bi-Laplacian of $m$
(see~\cite{FourdrinierStrawderman2003}). Thus the above domination
will occur as soon as
%
\begin{eqnarray} \label{equa12}
 \biggl(\frac{\Delta m (X)}{m(X)} \biggr)^2 - 2   \frac{\Delta^{(2)}
m (X)}{m (X)} \leq0   .
\end{eqnarray}
Applicability of that last condition is underlined by the remarkable
fact that if the prior
$\pi$ satisfies (\ref{equa12}), that is, if
%
\begin{eqnarray} \label{equa13}
 \biggl(\frac{\Delta\pi(\theta)}{\pi(\theta)} \biggr)^2 -
2   \frac{\Delta^{(2)} \pi(\theta)}{\pi(\theta)}
\leq0   ,
\end{eqnarray}
then (\ref{equa12}) is satisfied for the marginal $m$.

As an example, Fourdrinier and Strawderman \cite
{FourdrinierStrawderman2003} considered
$\pi(\theta) = (\| \theta\|^2 / 2 + a)^{-b}$
(where $a \geq0$ and $b \geq0$) and showed that, if $ p \geq2 (b + 3)$
then (\ref{equa13}) holds and hence $\delta_0$ dominates $\delta_m$.
Since $\pi$ is integrable if and only if $b > \frac{p}{2}$ (for $a >0$),
the prior $\pi$ is improper whenever this condition for
domination of~$\delta_0$ over $\delta_m$ holds. Of course, whenever
$\pi$ is proper, the Bayes estimator $\delta_m$ is admissible
provided its Bayes risk is finite.

Inadmissibility of the generalized Bayes loss estimator is not
exceptional. Thus, in
\cite{FourdrinierStrawderman2003}, the following general
inadmissibility result is given; its proof is parallel to the proof of
Theorem \ref{th2.1}.

\begin{theorem} \label{th5.1}
Let $m$ be a nonnegative function such that $\nabla m / m \in
W_{\mathrm{loc}}^{1,1}(\R^p)$.
Let $\xi$ be a real-valued function satisfying the conditions of
Theorem \ref{th2.1}.
Then $\delta_m$ is inadmissible and a class of dominating estimators
is given by
\[
\delta_m (X) + \alpha  \operatorname{sgn}   (\Delta\xi(X) )
\frac{\xi(X)}{m (X)} \quad
\mbox{for }    0 < \alpha< 2   K_0   .
\]
\end{theorem}

Note that, unlike Theorem \ref{th2.1}, neither the superharmonicity
condition nor the
subharmonicity condition on $m$ is needed. Note also that Theorem \ref
{th5.1} gives
conditions of improvement on $\delta_m$ while Theorem \ref{th2.1}
looks for improvements on
$\delta_0$. As we saw, often $\delta_0$ dominates $\delta_m$. So it
is not surprising
that the proofs of the two theorems are parallel; more precisely, it
suffices to suppress, in
the proof of Theorem \ref{th2.1}, the superharmonicity (or
subharmonicity) condition on $m$
to obtain the proof of Theorem~\ref{th5.1}.\looseness=1

In \cite{FourdrinierStrawderman2003}, it is suggested that the
inadmissibility of the generalized
Bayes (or pseudo-Bayes) estimator is due to the fact that the loss function
$(\delta(x) - \| \varphi(x) - \theta\|^2)^2$ may be inappropriate.
The possible deficiency
of this loss is illustrated by the following simple result concerning
estimation of the square
of a location parameter in $\R^1$.

Suppose $X \in\R\sim f  ( (X - \theta)^2  )$ such that
$E_{\theta}[ X^4] < \infty$.
Consider estimation of $\theta^2$ under loss $(\delta- \theta^2)^2$.
The generalized Bayes
estimator $\delta_{\pi}$ of $\theta^2$ with respect to the uniform prior
$\pi(\theta) \equiv1$ is given by
\[
\delta_{\pi} (X) = \frac{\int\theta^2 f  ( (X - \theta)^2
 )\,d\theta}
{\int f  ( (X - \theta)^2  )\,d\theta} = X^2 + E_0
[X^2]   .
\]
Since this estimator has constant bias $2   E_0[X^2]$, it is dominated
by the unbiased
estimator $X^2 - E_0[X^2]$ (the risk difference is $4
(E_0[X^2])^2$). Hence
$\delta_{\pi}$ is inadmissible for any $f (\cdot)$ such that
$E_{\theta}[ X^4] < \infty$.

\subsection{Examples of Improved Estimators} \label{subsection2.3}

In this subsection, we give some examples of Theorems \ref{th2.1} and
\ref{th5.1}.
The only example up to this point of an improved estimator over the
unbiased estimator of
loss $\delta_0(X)$ is $\delta(X) = \delta_0(X) + \alpha/ \|x\|^2$
for $0 < \alpha< 4   p$,
which is Johnstone's result \cite{Johnstone1988}. Although the
shrinkage factor in Theorems
\ref{th2.1} and \ref{th5.1} is the same, in the examples below we
will only focus on
improvements of posterior risk.

As an application of Theorem \ref{th5.1}, let
$\xi_b (x) = \break ( \| x \|^2 + a  )^{-b}$ (with $a \geq0$ and
$b \geq0$). It can be
shown that we have
$\Delta\xi_b (x) < 0$ for $a \geq0$ and $0 < 2   (b + \allowbreak 1) <
p$.\vadjust{\goodbreak}
Also $\Delta\xi_b (x) > 0$ if $a = 0$ and $2 (b + 1) > p$.
Furthermore
\begin{eqnarray*}
 \frac{\xi^2_b (x)}{| \Delta\xi_b (x) |} &=& \frac{1}{2   b
| p - 2
(b + 1) \fracc{\| x \|^2}{\| x \|^2 + a}  |} \\
&&{}\cdot  \frac{1}{(\| x \|
^2 + a)^{b - 1}}   .
\end{eqnarray*}

(a) Suppose that $0 < 2   (b + 1) < p$ and $a \geq0$. Then
\[
\frac{\xi^2_b (x)}{| \Delta\xi_b (x) |} \leq\frac{1}{2   b (p- 2
(b + 1))}
\frac{1}{(\| x \|^2 + a)^{b - 1}}
\]
and
$E_\theta [  (\Delta\xi_b(X)/\xi_b(X)  )^2  ]
< \infty$
since it is bounded from above by a quantity proportional to
$E_\theta [ ( \| X \|^2 + a )^{-2}  ]$,
which is finite for $a > 0$ or for $a = 0$ and $p > 4$.

Suppose that $m (x)$ is greater than or equal to some multiple of
$ ( \| x \|^2 + a  )^{1 - b} $ or equivalently
%
\begin{eqnarray}\label{equa2.26}
 \qquad m (x) \geq\frac{k}{2   b   (p - 2   (b + 1))}
\frac{1}{ ( \| x\|^2 + a  )^{b - 1}}
\end{eqnarray}
for some $k > 0$. Theorem \ref{th5.1} implies that $\delta_m (X)$ is
inadmissible and is
dominated by
\[
\delta_m (X) - \frac{\alpha}{m (X) (\| X \|^2 + a )^b}
\]
for $0 < \alpha< 4 b (p - 2 (b + 1)) \inf_{x \in\R^p}  ( m (x) (
\| x \|^2 +\break a)^{b - 1}  )$.
Note that the improved estimators shrink toward~$0$.

Suppose, for example, that $m (x) \equiv1$. Then (\ref{equa2.26}) is
satisfied for $b \geq1$.
Here $\varphi_m (X) = X$ and \mbox{$\delta_m (X) = p$}. Choosing $b = 1$, an
improved class of
estimators is given by $p - \frac{\alpha}{\| X \|^2 + a}$ for $0 <
\alpha< 4
(p - 4)$. The case $a = 0$ is equivalent to Johnstone's result for this
marginal.

(b) Suppose that $2(b + 1) > p > 4$ and $a = 0$. Then
\[
\frac{\xi^2_b (x)}{| \Delta\xi_b (x) |} = \frac{1}{2  b (2 (b +
1) - p)}
\frac{1}{\| x \|^{2 (b - 1)}}.
\]

A development similar to the above implies that, when $m (x)$ is greater
than or equal to some multiple of $\|x \|^{2(1-b)}$, an improved
estimator is
\[
\delta_m (X) + \frac{\alpha}{m (X) \| X \|^{2 b}}
\]
for $ 0 < \alpha< 4b (2   (b + 1) - p) \inf_{x \in\R^p}
 ( m (x) \| x \|^{2 (b - 1)}  ). $

Note that, in this case, the correction term is positive and hence
the estimators expand away from~$0$. Note also that this result
only works for $a = 0$ and hence applies to pseudo-marginals which
are unbounded in a neighborhood of $0$. Since all marginals
corresponding to a generalized prior $\pi$ are bounded, this result
can never apply to generalized Bayes procedures but only to
pseudo-Bayes procedures.

Suppose, for example, that $m (x) = \| x \|^{2-p}$.
Here $\varphi_m (X) =  ( 1 - \frac{p - 2}{\| x \|^2}  ) X$
is the
James--Stein estimator and $\delta_m (X) = p - \frac{(p - 2)^2}{\| X
\|^2}$.
In particular, the above applies for $b - 1 = \frac{p - 2}{2}$, that
is, for
$b = \frac{p}{2} > \frac{p - 2}{2}$. An improved estimator is given
by $\delta_m
(X) + \frac{\gamma}{\| X \|^2}$ for $0 < \gamma< 4   p$. This again agrees
with Johnstone's result for James--Stein estimators.


\section{Estimating the Quadratic Loss of a $p$-Normal Mean with
Unknown Variance}
\label{Section3}

In Section \ref{Section2} it was assumed that the covariance matrix
was known and equal to the identity matrix~$I_p$. Typically, the
covariance is unknown and should
be estimated. In the case where it is of the form $\sigma^2 I_p$ with
$\sigma^2$ unknown, Wan
and Zou \cite{WanZou2004} sho\-wed that, for the invariant loss
$\|\varphi(X) - \theta\|^2 / \sigma^2$,
Johnstone's result \cite{Johnstone1988} can be extended
when estimating the loss of the James--Stein estimator. In fact, the
general framework
considered in Section \ref{Section2} can be extended to the case where
$\sigma^2$ is unknown,
and we show that a~condition parallel to Condition~(\ref{domincond})
can be found.

Before stating the main result for the unknown variance case, we
need an extension of Stein's identity involving the sample variance.

\begin{lemma} \label{LemmaA1}
Let $X \sim{\cal N}(\theta,\sigma^2 I_p)$ and let $S$ be a~nonnegative random variable
independent of $X$ such that $S \sim\sigma^2 \chi_k^2$. Denoting by
$E_{\theta,\sigma^2}$ the
expectation with respect to the joint distribution of $(X,S)$, we have,
provided the
corresponding expectations exist, the following two results:
\begin{longlist}[(ii)]
\item[(i)] if $g(x,s)$ is a function from $\R^p \times\R_+$ into $\R^p$
such that, for any
$s \in\R_+$, $g(\cdot,s)$ is weakly differentiable, then
\[
E_{\theta,\sigma^2} \biggl[\frac{1}{\sigma^2}(X-\theta)^t
g(X,S) \biggr]
= E_{\theta,\sigma^2}[\operatorname{div}_X g(X,S)],
\]
where $\operatorname{div}_x g(x,s)$ is the divergence of $g(x,s)$ with
respect to $x$;

\item[(ii)] if $h(x,s)$ is a function from $\R^p \times\R_+$ into $\R$
such that, for any $s \in\R_+$,
$h(\cdot,s)$ is weakly differentiable, then
\begin{eqnarray*}
&&E_{\theta,\sigma^2} \biggl [\frac{1}{\sigma^2}   h(X,S) \biggr]\\
&& \quad  =
E_{\theta,\sigma^2}  \biggl[
2   \frac{\partial}{\partial S}h(X,S) + (k-2)S^{-1}   h(X,S)
 \biggr]
  .
\end{eqnarray*}
\end{longlist}
\end{lemma}
\begin{pf}
Part (i) is just Stein's lemma (cf. \cite{Stein1981}). Part (ii)
can\vadjust{\goodbreak}
be seen as a particular case
of Lem\-ma~1(ii) (established for elliptically symmetric distributions)
of Fourdrinier et al.
\cite{FourdrinierStrawdermanWells2003}, although we will present a~direct proof. The joint
distribution of $(X,S)$ can be viewed as resulting, in the
setting of the canonical
form of the general linear model, from the distribution of
$(X,U) \sim{\cal N}((\theta,0),\sigma^2 I_{p+k})$ with $S =
\|U\|^2$. Then we can write
\begin{eqnarray*}
&&E_{\theta,\sigma^2}  \biggl[\frac{1}{\sigma^2}   h(X,S) \biggr]\\
 && \quad =
E_{\theta,\sigma^2}  \biggl[\frac{1}{\sigma^2}   U^t   \frac
{U}{\|U\|^2}   h(X,\|U\|^2) \biggr]
\\
&& \quad =
E_{\theta,\sigma^2}  \biggl[\operatorname{div}_U \biggl(\frac
{U}{\|U\|^2}   h(X,\|U\|^2) \biggr) \biggr]
\end{eqnarray*}
according to part (i). Hence, expanding the divergence term, we have
\begin{eqnarray*}
&&E_{\theta,\sigma^2} \biggl [\frac{1}{\sigma^2}   h(X,S) \biggr]\\
 && \quad =
E_{\theta,\sigma^2}  \biggl[\frac{k-2}{\|U\|^2}   h(X,\|U\|^2)\\
&&\hphantom{\quad =
E_{\theta,\sigma^2}  \biggl[}{} +
\frac{U^t}{\|U\|^2}   \nabla_U h(X,\|U\|^2) \biggr]
\\
&& \quad =
E_{\theta,\sigma^2}  \biggl[\frac{k-2}{S}   h(X,S) +
2   \frac{\partial}{\partial S} h(X,S) \biggr]
\end{eqnarray*}
since
\[
\nabla_U h(X,\|U\|^2) = 2   \frac{\partial}{\partial S} h(X,S)
\bigg|_{S=\|U\|^2}   U   .
\]
\upqed
\end{pf}

The following theorem provides an extension of results in Section \ref
{Section2} to the setting
of an unknown variance. The necessary conditions to insure the
finiteness of the risks are given
in Appendix \ref{A.1}.

\begin{theorem} \label{unknownvar}
Let $X \sim{\cal N}(\theta,\sigma^2 I_p)$ where $\theta$\break and~$\sigma^2$ are unknown and
$p \geq5$ and let $S$ be a nonnegative random variable independent of
$X$ and such that
$S \sim\sigma^2 \chi_k^2$. Consider an estimator of $\theta$ of the form
$\varphi(X,S) = X + S   g(X,S)$ with $E_{\theta,\sigma^2}[S^2
\|g(X,\break S)\|^2] < \infty$,
where $E_{\theta,\sigma^2}$ denotes the expectation with respect to
the joint distribution of
$(X,S)$.

Then an unbiased estimator of the invariant loss $\|\varphi(X,S) -
\theta\|^2 / \sigma^2$ is
%
\begin{eqnarray}\label{unbiased2}
&&\hspace*{20pt}\delta_0(X,S)\nonumber\\
&&\hspace*{20pt}\quad{}= p + S  \biggl\{(k+2)   \|g(X,S)\|^2+ 2\operatorname{div}_Xg(X,S)\\
&&\hspace*{89pt}\hphantom{  = p + S  \biggl\{}{}+ 2   S \frac{\partial}{\partial S} \|g(X,S)\|^2 \biggr\}   .
\nonumber
\end{eqnarray}
Its risk
${\cal R}(\theta, \sigma^2, \varphi, \delta_0) =
E_{\theta,\sigma^2}[(\delta_0(X,S) - \|\varphi(X,\break S)-\theta\|^2 /
\sigma^2))^2]$
is finite as soon as $E_{\theta,\sigma^2}[S^2   \|g(X,\allowbreak S)\|^4] <
\infty$,
$E_{\theta,\sigma^2}[(S   \operatorname{div}_Xg(X,S))^2] < \infty$ and\break 
$E_{\theta,\sigma^2}     [ (S^2   \frac{\partial
}{\partial S}
\|g(X,S)\| )^2 ]  < \infty$.

Furthermore, for any function $\gamma(X)$ such that\break 
$E_{\theta,\sigma^2}[\gamma^2(X)] < \infty$,
the risk difference
${\cal D}(\theta,\sigma^2,\varphi,\allowbreak \delta) = {\cal R}(\theta,
\sigma^2, \varphi, \delta)
- {\cal R}(\theta, \sigma^2, \varphi, \delta_0)$
between the estimators
$\delta(X,S) = \delta_0(X,S) - S   \gamma(X)$ and $\delta_0(X,S)$
is given by
%
\begin{eqnarray}\label{unbiasedrisk2}
&&\hspace*{-6pt}E_{\theta,\sigma^2}
 \biggl[
S^2   \biggl \{  \gamma^2(X) + \frac{2}{k+2}   \Delta\gamma(X)
\nonumber
\\[-6pt]
\\[-8pt]
&&\hspace*{38.5pt}{}+ 4   g^t(X,S) \nabla  \gamma(X) + 4   \gamma(X)   \|g(X,S)\|^2
   \biggr\}
 \biggr].
\nonumber
\end{eqnarray}
Therefore a sufficient condition for ${\cal D}(\theta,\sigma
^2,\varphi,\delta)$ to be nonpositive, and hence for $\delta(X,S)$
to improve on $\delta_0(X,S)$, is
%
\begin{eqnarray} \label{domincond2}
&&\gamma^2(x) + \frac{2}{k+2}   \Delta\gamma(x)
+ 4   g^t(x,s)   \nabla\gamma(x) \nonumber
\nonumber
\\[-8pt]
\\[-8pt]&& \quad {}+ 4   \gamma(x)   \|g(x,s)\|^2
\leq0
\nonumber
\end{eqnarray}
for any $x \in\R^p$ and any $s \in\R_+$.
\end{theorem}
\begin{pf}
According to the expression of $\varphi(X,S)$, its risk $R(\theta
,\varphi)$ is the expectation
of
%
\begin{eqnarray}\label{risk7}
&&\frac{1}{\sigma^2}   \|X-\theta\|^2 + 2   \frac{S}{\sigma^2}
(X-\theta)^tg(X,S)
\nonumber
\\[-8pt]
\\[-8pt]&& \quad {}+ \frac{S^2}{\sigma^2}   \|g(X,S)\|^2   .
\nonumber
\end{eqnarray}
Clearly $
E_{\theta,\sigma^{2}} [\sigma^{-2}   \|X-\theta\|^2 ] =
p $ and Lemma \ref{LemmaA1} implies that
\[
E_{\theta,\sigma^2} \biggl[\frac{1}{\sigma^2}(X-\theta)^t
g(X,S) \biggr]
= E_{\theta,\sigma^2}[\operatorname{div}_X g(X,S)]
\]
and, with $h(x,s) = s^2   \|g(x,s)\|^2$, that
\begin{eqnarray*}
&&E_{\theta,\sigma^2}  \biggl[\frac{S^2}{\sigma^2}
\|g(X,S)\|^2 \biggr]\\
&& \quad  =
E_{\theta,\sigma^2}  \biggl[S \biggl \{(k+2)   \|g(X,S)\|^2 \\
&& \hphantom{\quad  =
E_{\theta,\sigma^2}  \biggl[S \biggl \{}  {}+
2   S   \frac{\partial}{\partial S}\|g(X,S)\|^2 \biggr\} \biggr]
  .
\end{eqnarray*}
Therefore $R(\theta,\varphi) = E_{\theta,\sigma^2}[\delta_0(X,S)]$
with $\delta_0(X,S)$ gi\-ven
in (\ref{unbiased2}), which means that $\delta_0(X,S)$ is an unbiased
estimator of the invariant loss
$\|\varphi(X,S) - \theta\|^2 / \sigma^2$. The fact that the risk
${\cal R}(\theta, \sigma^2, \varphi,
\delta_0)$ of $\delta_0(X)$ is
finite is shown in Lemma \ref{LemmaA2}.

Now consider the finiteness of the risk of the alternative loss
estimator
$\delta(X,S) = \delta_0(X,S) - S   \gamma(X)$. It is easily seen\vadjust{\goodbreak}
that its difference in loss
$d(\theta,\sigma^2,X , S)$ with $\delta_0(X,S)$ can be written as
%
\begin{eqnarray}\label{loss7}
&&d(\theta,\sigma^2, X, S)
        \nonumber\\  && \quad =           \biggl (
    \delta_0(X,S)   -
\frac{1}{\sigma^2} \|\varphi(X)   -   \theta\|^2 - S   \gamma(X)
 \biggr)^{    2}\nonumber\\
 && \qquad {}
-    \biggl(    \delta_0(X,S) - \frac{1}{\sigma^2} \|\varphi(X)
  -   \theta\|^2
 \biggr)^{    2} \\
&& \quad = S^2   \gamma^2(X)\nonumber\\
&& \qquad {} - 2   S   \gamma(X)
\biggl(\delta_0(X,S) - \frac{1}{\sigma^2}\|\varphi(X)-\theta\|^2\biggr)
  .
\nonumber
\end{eqnarray}
Hence, since $E_{\theta,\sigma^2}[\|\varphi(X,S) - \theta\|^2 /
\sigma^2] < \infty$ as the risk
of the estimator $\varphi(X,S)$, the condition\break 
$E_{\theta,\sigma^2}[\gamma^2(X)] < \infty$
ensures that the expectation of the loss in (\ref{loss7}), that is,
the risk difference
${\cal D}(\theta,\sigma^2,\break\varphi,\delta)$ is finite. Then
${\cal R}(\theta, \sigma^2, \varphi, \delta) < \infty$ since
${\cal R}(\theta, \sigma^2,\allowbreak \varphi,  \delta_0) < \infty$.

We now express the risk difference
${\cal D}(\theta,\sigma^2,\varphi,\delta) =
E_{\theta,\sigma^2}[d(\theta,\sigma^2,X , S)]$. Using (\ref
{unbiased2}) and expanding\break 
$\|\varphi(X,S) - \theta\|^2 / \sigma^2$ give that $d(\theta,\sigma
^2,X , S)$ in (\ref{loss7})
can be written as
$ d(\theta,\sigma^2,X , S) = A(X,S) + B(\theta, \sigma^2,\allowbreak  X, S) $
where
%
\begin{eqnarray}\label{losspart7a}
A(X,S) &=& S^2   \gamma^2(X)
- 2   p  S   \gamma(X) \nonumber\\&&{}- 2   (k+2)   S^2   \gamma(X)
\|g(X,S)\|^2 \nonumber
\\[-8pt]
\\[-8pt]
&&{}- 4   S^2   \gamma(X)   \operatorname{div}_Xg(X,S)\nonumber\\
&&{}-
4   S^3   \gamma(X)   \frac{\partial}{\partial S} \|g(X,S)\|^2
\nonumber
\end{eqnarray}
and
%
\begin{eqnarray}\label{losspart7b}
B(\theta, \sigma^2, X, S) &=& 2   \frac{S^3}{\sigma^2}   \gamma(X)
  \|g(X,S)\|^2
   \nonumber\\
   &&{}  +
    2   \frac{S}{\sigma^2}   \gamma(X)   \|X-\theta\|^2
\\
    &&{}+     4   \frac{S^2}{\sigma^2}   \gamma(X)   (X-\theta
)^tg(X,S)   .
\nonumber
\end{eqnarray}

Through Lemma \ref{LemmaA1}(ii) with
$h(x,s) = 2   \frac{s^3}{\sigma^2}   \gamma(x) \*  \|g(x,s)\|^2$,
the expectation of the
first term in the right-hand side of (\ref{losspart7b}) equals
%
\begin{eqnarray}\label{losspart7b1}
&&E_{\theta,\sigma^2}
 \biggl[
2   \frac{S^3}{\sigma^2}   \gamma(X)   \|g(X,S)\|^2
 \biggr]\nonumber\\
&& \quad =
E_{\theta,\sigma^2}      \biggl[ 2   (k+4)   S^2   \gamma(X)
\|g(X,S)\|^2\\
&&  \hspace*{4.5pt}\hphantom{\quad =
E_{\theta,\sigma^2}      \biggl[} {} +
4   S^3   \gamma(X)   \frac{\partial}{\partial S}\|g(X,S)\|^2
 \biggr]
.
\nonumber
\end{eqnarray}

An iterated application of Lemma \ref{LemmaA1}(i) to the expectation
of the second term\vadjust{\goodbreak}
in the right-hand side of (\ref{losspart7b}) allows to write
\begin{eqnarray*}
&&E_{\theta,\sigma^2}  \biggl[2   \frac{S}{\sigma^2}   \gamma(X)
\|X-\theta\|^2  \biggr]\\[-1pt]
&& \quad =
E_{\theta,\sigma^2}\biggl[2   \frac{1}{\sigma^2}   (X-\theta)^t   S
  \gamma(X)   (X-\theta)\biggr]
\nonumber\\[-1pt]
&& \quad =
E_{\theta,\sigma^2}[2  \operatorname{div}_X \{S   \gamma(X)
(X-\theta)\}] \nonumber\\[-1pt]
&& \quad =
E_{\theta,\sigma^2}[2   p   S   \gamma(X) + 2   S   (X-\theta
)^t \nabla\gamma(X)]
\nonumber\\[-1pt]
&& \quad =
E_{\theta,\sigma^2}[2   p   S   \gamma(X) + 2   \sigma^2   S
  \Delta\gamma(X)]
\end{eqnarray*}
which, as $S \sim\sigma^2 \chi_k^2$ entails that $E[S^2 / (k+2)] =
E[\sigma^2   S]$ and as
$S$ is independent of $X$, gives
%
\begin{eqnarray}\label{losspart7b2}
&&E_{\theta,\sigma^2} \biggl[2   \frac{S}{\sigma^2}   \gamma(X)
\|X-\theta\|^2 \biggr]\nonumber
\\[-9pt]
\\[-9pt]
&& \quad =
E_{\theta,\sigma^2} \biggl[2   p   S   \gamma(X) +
2   \frac{S^2}{k+2}   \Delta\gamma(X) \biggr]   .
\nonumber
\end{eqnarray}
As for the third term in the right-hand side of (\ref{losspart7b}),
its expectation can also
be expressed using Lem\-ma~\ref{LemmaA1}(i) as
%
\begin{eqnarray}\label{losspart7b3}
&&E_{\theta,\sigma^2}
 \biggl[4   \frac{S^2}{\sigma^2}   \gamma(X)   (X-\theta
)^tg(X,S) \biggr]\nonumber\\[-1pt]
&& \quad =
E_{\theta,\sigma^2}[4   S^2   \operatorname{div}_X \{\gamma(X)
g(X,S)\}] \nonumber
\\[-9pt]
\\[-9pt]
&& \quad =
E_{\theta,\sigma^2}[4   S^2   \gamma(X)   \operatorname{div}_X \{
g(X,S)\}\nonumber\\[-1pt]
&&\hspace*{5.5pt}\hphantom{\quad =
E_{\theta,\sigma^2}[}{} +
4   S^2   g(X,S)^t   \nabla\gamma(X)]
\nonumber
\end{eqnarray}
by the product rule for the divergence. Finally, gathering
(\ref{losspart7b1}), (\ref{losspart7b2}) and (\ref{losspart7b3})
yields an
expression of~(\ref{losspart7b}) which, with (\ref{losspart7a}),
gives the integrand term of~(\ref{unbiasedrisk2}), which is the desired result.
\end{pf}

As an example, consider the James--Stein estimator with unknown variance
\[
\varphi^{\mathrm{JS}}(X,S) = X - \frac{p-2}{k+2} \frac{S}{\|X\|^2} X   .
\]
Here the shrinkage factor is the product of a function of $S$ with a
function of $X$
so that, through routine calculation, the unbiased estimator of loss is
\[
\delta_0(X,S) = p - \frac{(p-2)^2}{k+2} \frac{S}{\|X\|^2}   .
\]
For a correction of the form $\gamma(x) = - d/\|x\|^2$ with $d \geq
0$, it is easy to
check that the expression in (\ref{domincond2}) equals
\begin{eqnarray*}
&&d^2 + 4   \frac{p-4}{k+2}   d - 8   \frac{p-2}{k+2} \, d
- 4    \biggl(\frac{p-2}{k+2} \biggr)^2 \, d\\[-1pt]
&& \quad
=
d \biggl (d - \frac{4}{k+2} \biggl [p + \frac{(p-2)^2}{k+2}
\biggr] \biggr)
\end{eqnarray*}
which is negative for $0 < d < \frac{4}{k+2}  [p + \frac
{(p-2)^2}{k+2} ]$ and gives
domination of $p - \frac{(p-2)^2}{k+2} \frac{S}{\|X\|^2} + \frac
{d}{\|x\|^2}$ over
$p - \frac{(p-2)^2}{k+2} \frac{S}{\|X\|^2}$. This condition recovers
the result of Wan and Zou
\cite{WanZou2004} who considered the case
$d = \frac{2}{k+2}  [p + \frac{(p-2)^2}{k+2} ]$.


\section{Extensions to the Spherical Case} \label{Section4}

\subsection{Estimating the Quadratic Loss of the Mean of a Spherical
Distribution}
\label{subsection4.1}

In the previous sections the loss estimation problem was considered
for the normal distribution setting. The normal distribution has
been generalized in two important directions, first as a special
case of the exponential family and second as a spherically
symmetric distribution. In this section we will consider the
latter. There are a variety of equivalent definitions and
characterizations of the class of spherically symmetric
distributions; a comprehensive review is given in \cite{FangKotzNg1990}.
We will use the representation of a random variable from a
spherically symmetric distribution, $X = (X_{1}, \ldots,X_{p})^t$,
as $X \stackrel{d}{=} RU^{(p)} + \theta$, where $R= \| X -\theta\|$
is a random radius, $U^{(p)}$ is a~uniform random variable on the
$p$-dimensional unit sphe\-re, where $R$ and $U^{(p)}$ are
independent. In such a~situation, the distribution of $X$ is said to be
spherically symmetric around $\theta$ and we write $X \sim
\operatorname{SS}_{p}(\theta)$. We also extend, in  Section \ref{subsection4.2},
these results to the case where the distribution of $X$ is
spherically symmetric and when a residual vector $U$ is available
(which allows an estimation of the variance factor~$\sigma^2$).

Assume $X \sim \operatorname{SS}_{p}(\theta)$ and suppose we wish to estimate
$\theta\in\R^{p}$ by a decision rule $\delta(X)$ using quadratic
loss. Suppose that we also use quadratic loss to assess the accuracy
of loss estimate $\delta(X)$; then the risk of this loss estimate
is given by (\ref{newrisk}). In \cite{FourdrinierWells1995b}, the
problem of estimating the loss when $\varphi(X) = X$ is the
estimate of the location parameter $\theta$ is considered. The
estimate $\varphi$
is the least squares estimator and is minimax among the class of
spherically symmetric distributions with bounded second moment.
Furthermore, if one assumes the density of $X$ exists and is
unimodal, then $\varphi$ is also the maximum likelihood estimator.

The unbiased constant estimate of the loss $\| X - \theta\|^{2}$ is
$\delta_{0} = E_{\theta}[R^{2}]$. Note that $\delta_{0}$ is independent
of~$\theta$, since $E_{\theta}[\| X - \theta\|^{2}] = E_{0}[\| X \|^{2}]$.
Fourdrinier and\break Wells~\cite{FourdrinierWells1995b} showed that the unbiased estimator
$\delta_{0}$ can be dominated by $\delta_{0} - \gamma$, where
$\gamma$ is a particular superharmonic function for the case
where\vadjust{\goodbreak}
the sampling distribution is a scale mixture of normals and in more
general spherical cases.

The development of the results depends on some interesting
extensions of the classical Stein identities in
(\ref{SteinIdentity}) and (\ref{unbiasednorme22}) to the general
spherical setting. Since the distribution of $X$, say $P_\theta$,
is spherically symmetric around $\theta$, for every bounded function
\textit{f}, we have $ E_{\theta}[f] = E   E_{R, \theta}[f] =
\int_{\R_{+}} E_{R, \theta}[f]\rho(dR), $ whe\-re~$\rho$ is the
distribution of the radius, namely the distribution of the norm $\|X
- \theta\|$ under $P_\theta$ and where~$E$ and $E_{R, \theta}$ denote
respectively the expectation with respect to the radial distribution
and uniform distri\-bution $U_{R, \theta}$ on the sphere $S_{R,
\theta} = \{x \in\R^{p} \mid  \| x - \theta\| = R\}$ of radius $R$
and center $\theta$. To deduce the various risk domination results
it suffices to work conditionally on the radius, that is to say to
replace $P_{\theta}$ by~$U_{R, \theta}$ in the risk expressions. Let
$\sigma_{R, \theta}$ denote the area measure on $S_{R, \theta}$.
Therefore, for every Borel measurable set $A$, $ U_{R, \theta}(A) =
\sigma_{R, \theta}(A)/\sigma(S_{R, \theta}) = \Gamma(p/\break2) \sigma_{R,
\theta}(A)/2 \pi^{p/2}R^{p-1}. $ Define the volume measu-\break re~$\tau_{R,
\theta}$ on the ball $B_{R, \theta} = \{x \in\R^p \mid  \| x -
\theta\|
\leq R \}$~of radius $R$ and center $\theta$ and denote the uniform
distri\-bution on $B_{R, \theta}$ as $V_{R, \theta}$. Hence, for
every Borel measurable set $A$, $ V_{R, \theta}(A) = \tau_{R,
\theta}(A)/\tau_{R, \theta} (B_{R, \theta}) = p \Gamma(p/\allowbreak2)\tau_{R,
\theta}(A)/2 \pi^{p/2}R^{p}.$ Suppose $\gamma$ is a weakly
differentiable vector-valued function; then by applying the
Divergence Theorem for weakly differentiable functions to the
definition of the expectation we have
%
\begin{eqnarray} \label{ssstein1}
&&E_{\theta}[(X - \theta)^t \gamma(X) \mid\| X - \theta\| =
R]\nonumber\\
&& \quad =
\int_{S_{R, \theta}} (x - \theta)^{t}   \gamma(x)   U_{R, \theta
}(dx) \\
&& \quad =
\frac{R}{\sigma_{R, \theta}(S_{R, \theta})} \int_{B_{R,\theta}}
\operatorname{div}   \gamma(x) \, dx.
\nonumber
\end{eqnarray}
If $\gamma$ is a real-valued function, then it follows\break from~(\ref{ssstein1}) and the product rule applied to the vector-valued
function $(x - \theta)   \gamma(x)$ that
%
\begin{eqnarray} \label{ssstein2}
&&E_{\theta}[ \| X - \theta\|^{2} \gamma(X) \mid\| X - \theta\| =
R]\nonumber\\
&& \quad = \int_{S_{R, \theta}} (x - \theta)^{t}   (x - \theta)   \gamma
(x)   U_{R, \theta}(dx)\nonumber
\\[-8pt]
\\[-8pt]
&& \quad = \frac{R}{\sigma_{R, \theta}(S_{R, \theta})}\nonumber\\
&& \qquad {}\cdot \int_{B_{R, \theta}}
 [p   \gamma(x) + (x - \theta)^{t}   \nabla\gamma(x) ]\,dx   .
\nonumber
\end{eqnarray}

Our first extension of Theorem \ref{th2.1} is to the class of
spherically symmetric distributions that are scale mixtures of
normal distributions. Well-known examples in the class of densities
include the double exponential, multivariate $t$-distribution
(hence, the multivariate Cauchy distribution). Let $\phi(x; \theta,
I)$ be the probability
density function of a random vector~$X$ with a normal distribution with mean
vector~$\theta$ and identity covariance matrix. Suppose that there is
a~probability
measure on $\R_{+}$ such that the probability density function
$p_\theta$ may be expressed as
%
\begin{equation} \label{mixturepdf}
p_{\theta}(x| \theta) = \int_{0}^{\infty} \phi(x; \theta,
I/\varsigma)  G(d\varsigma).
\end{equation}

One can think of $\Upsilon$ being a random variable with distribution $G$;
the conditional distribution of $X$ given $\Upsilon= \varsigma, X
|\Upsilon= \varsigma$, is
$N_{p}(\theta, I/\varsigma)$. This class contains some
heavy-tailed distributions, possibly with no moments. It is well
known (see \cite{FangKotzNg1990}) that, if
a~sphe\-rical distribution has a density $p_{\theta}$, it is of the form
$p_{\theta}(x) = g(\| x - \theta
\|^{2})$ for a measurable positive function $g$ (called the
generating function).

In the scale mixture of normals setting the unbiased estimate,
$\delta_0$, of risk equals
\begin{eqnarray*}
E[R^2] = E_{\theta} [\|X-\theta\|^2] = p\int_{0}^{\infty} \varsigma
^{-1} G(d\varsigma).
\end{eqnarray*}

It is easy to see that the risk of the unbiased estimator
$\delta_0$ is finite if and only if $E_{\theta}
[\|X-\theta\|^4]<\infty$, which holds if
%
\begin{eqnarray} \label{mixcondition}
\int_{0}^{\infty} \varsigma^{-2} G(d\varsigma)<\infty.
\end{eqnarray}

The main theorem in \cite{FourdrinierWells1995b} is the following
domination result of an improved estimator of loss over the unbiased
loss estimator.
%
\begin{theorem}\label{th3.1}
Assume the distribution of $X$ is a scale mixture of normal random
variables as\break in~(\ref{mixturepdf}) such that (\ref{mixcondition}) is
satisfied and such that
%
\begin{equation} \label{mixsuff1}
\int_{\R_{+}} \varsigma^{p/2}   G(d\varsigma) < \infty  .
\end{equation}
Also, assume that the shrinkage function $\gamma$ is twice weakly
differentiable on $\R^{p}$ and satisfies $E_{\theta}[\gamma^{2}]
<\infty$,
for every $\theta\in\R^{p}$.
Then a sufficient condition for \mbox{$\delta_0 - \gamma$} to dominate
$\delta_0$ is that $\gamma$ satisfies the differential inequality
%
\begin{equation} \label{mixsuff2}
  k   \Delta\gamma+ \gamma^{2} < 0
\quad\mbox{with }
k = 2   \frac
{\int_{\R_{+}} \varsigma^{p/2}   G(d\varsigma)}{\int_{\R_{+}}
\varsigma^{p/2-2}   G(d\varsigma)}
  .\hspace*{-30pt}
\end{equation}
\end{theorem}

As an example let $\gamma(x) = c/\|x\|^{2}$ where $c$ is a positive
constant. Note that $\gamma$ is twice weakly differentiable only when
$p > 4$ (thus its
Laplacian exists as a locally integrable function). Then it may be shown
that $\Delta\gamma(x) = -2c(p - 4)/\| x \|^{4}$. Hence $k
\Delta(x) +\break \gamma^{2}(x) = -2kc(p-4)/\| x \|^{4} + c^{2}/\| x
\|^{4} < 0$ if $-2kc(p-4) + c^{2} < 0$, that is, $0 < c < 2k(p-4)$.
It is easy to see that the optimal value of $c$ for which this
inequality is the most negative equals $k(p - 4)$, so an interesting
estimate in this class of $\gamma$'s is $\delta= \delta_0 - k(p -
4)/ \|x\|^{2}$ $(p > 4)$. This is precisely the estimate proposed by~\cite{Johnstone1988} in the normal distribution case $N_{p}(\theta,
I)$ where $k = 2$; recall, in that case $\delta_0 = p$.

In this example, we have assumed that the dimension $p$ is greater than 4.
In general we can have domination as long as the assumptions of the
theorem are valid. Actually, Blanchard and Fourdrinier~\cite{BlanchardFourdrinier1999}
showed explicitly that, when $p \leq
4$, the only solution~$\gamma$ in $L^{2}_{\rm{loc}}(\R^{p})$ of the
inequality \mbox{$k \Delta\gamma+ \gamma^{2} \leq0$} is $\gamma\equiv
0$, almost everywhere with respect to the Lebes\-gue measure $\lambda$.
Now, in the normal setting $N_{p}(\theta, I/\varsigma)$, an unbiased
estimator of
the risk difference between an estimator $\delta= \delta_0 -\gamma$
and $\delta_0$ is $2\varsigma^{-2} \Delta\gamma+ \gamma^{2}$. Hence,
for dimensions 4 or less, it is impossible to find an estimator
$\delta= \delta_0 -\gamma$ whose unbiased estimate of risk is always
less than that of~$\delta_0$. Indeed we cannot have
$E_{\theta}[2\varsigma^{-2} \Delta\gamma+\gamma^{2}] < 0$,
for some $\theta$, without having
$\lambda[ \varsigma^{-2} \Delta\gamma(x) +\gamma^{2}(x) < 0 ] > 0$,
which entails that
$\lambda[ \gamma(x) \not= 0 ] > 0$.

In the case of scale mixture of normal distributions, the conjecture
of admissibility of $\delta_0$ for lower dimensions, although it is
probably true, remains open. Indeed, under conditions of Theorem
\ref{th3.1},\break $k \Delta\gamma+ \gamma^{2}$ is no longer an unbiased
estimator of the risk difference and $E_{\theta}[k \Delta\gamma+
\gamma^{2}]$ is only its upper bound. The use of Blyth's method
would need to specify the distribution of $X$ (i.e., the mixture
distribution $G$). It is worth noting that dimension-cutoff also
arises through the finiteness of $E_{\theta}[\gamma^{2}]$ when using
the classical shrinkage function $c/\|x \|^{2}$.

In order to prove Theorem \ref{th3.1} we need some additional
technical results. The first lemma gives some important properties of
superharmonic
functions and is found in Du Plessis \cite{Duplessis1970} and the second
lemma links the integral of the gradient on a ball with the integral
of the Laplacian.

\begin{lemma}\label{lemma3.2}
If $\gamma$ is a real-valued superharmonic function, then:
\begin{longlist}[(ii)]
\item[(i)] $\int_{S_{R, \theta}} \gamma(x) U_{R, \theta}(dx) \leq\int_{B_{R,
\theta}} \gamma(x) V_{R, \theta}(dx)$,

\item[(ii)] both of the integrals in \textup{(i)} are decreasing in~$R$.
\end{longlist}
\end{lemma}

\begin{pf} See Sections 1.3 and 2.5 in \cite{Duplessis1970}.
\end{pf}

\begin{lemma}\label{lemma3.3}
Suppose $\gamma$ is a twice weakly differentiable function. Then
\begin{eqnarray*}
&&\int_{B_{R, \theta}} (x - \theta)^{t} \nabla\gamma(x) V_{R,
\theta}(dx)\\
&& \quad  = \frac{p \Gamma(p/2)}{2 \pi^{p/2}} \frac{1}{R^{p}}
\int^{R}_{0} r \int_{B_{r, \theta}} \bigtriangleup\gamma(x)\,dx\, dr.
\end{eqnarray*}
\end{lemma}
\begin{pf} Since the density of the distribution of the radius
under $V_{R, \theta}$ is $(p/R^{p}) r^{p-1} \nbOne_{[0,R]}(r)$, we have
\begin{eqnarray*}
&&\int_{B_{R, \theta}} (x - \theta)^t \nabla\gamma(x) V_{R,
\theta}(dx)\\
&& \quad = \int^{R}_{0} \int_{S_{r, \theta}} (x - \theta)^{t}
\nabla\gamma(x) U_{r, \theta}(dx)\frac{p}{R^{p}} r^{p-1}\,dr.
\end{eqnarray*}
The result follows from applying (\ref{ssstein1})
to the innermost integral of the right-hand side of this equality
and by recalling the fact that $\sigma_{r, \theta}(S_{r, \theta}) =
(2 \pi^{p/2}/\break \Gamma(p/2))\* r^{p-1}$.
\end{pf}

\begin{pf*}{Proof of Theorem \ref{th3.1}}
Denoting by $\rho$ the
distribution of the
radius $\|X - \theta\|$, the risk difference between $\delta_0$ and
$\delta_{0}-\gamma$
equals $\alpha(\theta) + \beta(\theta)$ where
%
\begin{eqnarray}\label{alphabeta}
\alpha(\theta) &=& \int_{\R_{+}} \alpha_R(\theta)   \rho(dR)
\quad\mbox{and}  \nonumber
\\[-8pt]
\\[-8pt]
 \beta(\theta) &=& \int_{\R_{+}} \beta_R(\theta)   \rho(dR)
\nonumber
\end{eqnarray}
with
%
\begin{eqnarray}\label{alphaR}
\alpha_R(\theta) &=& 2   R^2 \int_{B_{R, \theta}} \gamma(x)
V_{R,\theta}(dx)\nonumber
\\[-8pt]
\\[-8pt]
&&{}- 2   \lambda_0 \int_{S_{R, \theta}} \gamma(x)   U_{R, \theta}(dx)
\nonumber
\end{eqnarray}
and
%
\begin{eqnarray}\label{betaR}
\beta_R(\theta) &=& 2   \frac{R^{2}}{p}
\int_{B_{R, \theta}} (x - \theta)^{t}   \nabla\gamma(x)   V_{R,
\theta}(dx) \nonumber
\\[-8pt]
\\[-8pt]&&{}+
\int_{S_{R, \theta}} \gamma^2(x)   U_{R, \theta}(dx)   .
\nonumber
\end{eqnarray}
Indeed, the risk difference conditional on the radius~$R$ equals
\begin{eqnarray*}
\int_{S_{R, \theta}}
 [2   \|x - \theta\|^2   \gamma(x) - 2   \lambda_0   \gamma
(x) + \gamma^2(x) ]
U_{R, \theta}(dx)
\end{eqnarray*}
and the result follows from (\ref{ssstein2}) applied to the first term
between brackets.

Let us first deal with $\alpha(\theta)$ considering the first term in
(\ref{alphaR}).
We have from the definition of $V_{R, \theta}$ and an application of
Fubini's theorem
%
\begin{eqnarray} \label{prthm3.1_1}
&& \quad \int_{\R_{+}} R^{2} \int_{B_{R, \theta}} \gamma(x)   V_{R,\theta
}(dx)   \rho(dR)\hspace*{-12pt}\nonumber\\
    && \quad  \quad  =
p   \frac{\Gamma(p/2)}{2 \pi^{p/2}} \int_{\R_{+}} R^{2-p}
\int_{B_{R, \theta}} \gamma(x) \, dx   \rho(dR)\hspace*{-12pt} \\
    && \quad  \quad  =
p   \frac{\Gamma(p/2)}{2 \pi^{p/2}} \int_{\R^{p}} \gamma(x)
\int^{+ \infty}_{\| x - \theta\|} R^{2-p}   \rho(dR) \, dx.\hspace*{-12pt}
\nonumber
\end{eqnarray}
Now, for fixed $\varsigma\geq0$, in the normal case $N_{p}(\theta,
I/\varsigma)$
the distribution $\rho_{\varsigma}$ of the radius has the density
$f_{\varsigma}$ of the form
$ f_{\varsigma}(R) = \frac{\varsigma^{p/2}}{2^{p/2-1} \Gamma(p/2)}
R^{p-1} \exp\{- \frac{\varsigma R^{2}}{2}\} $
and $ \delta_0 = \frac{p}{\varsigma}$. Thus the expression (\ref
{prthm3.1_1}) becomes
\begin{eqnarray*} \label{prthm3.1_2}
&&\int_{\R_{+}} R^{2} \int_{B_{R, \theta}} \gamma(x)   V_{R,\theta
}(dx)   \rho(dR)\\
&& \quad  =
\frac{p   \varsigma^{p/2}}{(2 \pi)^{p/2}} \int_{\R^{p}} \gamma(x)
\int^{+ \infty}_{\| x - \theta\|} R \exp \biggl\{ - \frac{\varsigma
R^{2}}{2}  \biggr\}\,dR \,dx \\
&& \quad  =
\frac{p   \varsigma^{p/2-1}}{(2 \pi)^{p/2}}
\int_{\R^{p}} \gamma(x) \exp \biggl\{ - \frac{\varsigma}{2} \| x -
\theta\|^{2}  \biggr\}\,dx \\
&& \quad  =
\frac{p}{\varsigma} \int_{\R_{+}} \int_{S_{R, \theta}} \gamma(x)
  U_{R,\theta}(dx)   \rho_{\varsigma}(dR),
\end{eqnarray*}
the last equality holding since $X \stackrel{D}{=}
RU^{(p)}$. Turning back to (\ref{alphabeta}) and (\ref{alphaR}) and
using the mixture
representation with mixing distribution $G$, the expression of $\alpha
(\theta)$ is written as
%
\begin{eqnarray}\label{alpha(theta)}
\alpha(\theta) &=& 2 p \int_{\R_{+}}  \biggl( \frac{1}{\varsigma} -
\frac{\delta_0}{p}
 \biggr)\nonumber\\
 &&\hphantom{2 p \int_{\R_{+}}}{}\cdot \int_{\R^{p}} \gamma(x) \biggl ( \frac{\varsigma}{2 \pi}
 \biggr)^{p/2}
\\&&\hspace*{56pt}{}\cdot\exp\biggl ( - \frac{\varsigma}{2} \|x - \theta\|^{2}  \biggr)\,dx   G(d\varsigma).
\nonumber
\end{eqnarray}
It can be easily seen that the innermost integral in~(\ref
{alpha(theta)}) is proportional to
\begin{eqnarray*}
\int_0^\infty\int_{S_{(u/\varsigma)^{1/2}, \theta}} \gamma(x) \,
dU_{S_{(u/\varsigma)^{1/2}, \theta}}
u^{p/2-1}   \exp \biggl(-\frac{u}{2} \biggr)\,du
\end{eqnarray*}
and hence is nondecreasing in $\varsigma$ by superharmonicity of
$\gamma$ induced by the inequality in
(\ref{mixsuff2}) and by Lemma~\ref{lemma3.2}(ii). Thus, since $
\delta_0 = p / \varsigma$ for fixed
$\varsigma$, the expression for $\alpha(\theta)$ in (\ref
{alpha(theta)}) is a nonpositive
covariance with respect to $G$.

We can now treat the integral of the expres-\break sion~$\beta(\theta)$ in
the same manner.\vadjust{\goodbreak}
The function $x \rightarrow(x - \theta)^{t} \nabla\gamma(x)$ and
the function
$x \rightarrow\nabla\gamma(x)$ taking successively the role of the function
$\gamma$, we obtain
\begin{eqnarray*}
&&\int_{\R_{+}} \frac{R^{2}}{p} \int_{B_{R, \theta}} (x - \theta
)^{t} \nabla\gamma(x) V_{R, \theta}(dx) \rho_{\varsigma}(dR)
\\
&& \quad  = \frac{1}{\varsigma} \int_{\R_{+}} \int_{S_{R, \theta}} (x -
\theta)^{t} \nabla\gamma(x) U_{R, \theta}(dx) \rho_{\varsigma
}(dR)\\
&& \quad  =  \frac{1}{\varsigma} \int_{\R_{+}} \frac{R^{2}}{p} \int
_{B_{R, \theta}} \nabla\gamma(x)\,dx \rho_{\varsigma}(dR)\\
&& \quad  = \frac{\varsigma^{p/2-2}}{(2 \pi)^{p/2}} \int_{\R^{p}} \nabla
\gamma(x)
\exp \biggl\{ - \frac{\varsigma}{2} \| x - \theta
\|^{2}  \biggr\}\,dx
\end{eqnarray*}
by applying (\ref{ssstein1}) for the second equality and remembering that
$\triangle\gamma= \operatorname{div}(\nabla\gamma)$. Therefore by
the Fubini Theorem
$\beta(\theta)$ can be reexpressed as
%
{\fontsize{10.6pt}{12.6pt}\selectfont{
\begin{eqnarray}\label{betatheta}
\beta(\theta) & = & \int_{\R^{p}}  \biggl( 2 \bigtriangleup\gamma
(x) \nonumber\\
&&\hphantom{\int_{\R^{p}}  \biggl(}{}\cdot\frac{\int_{\R_{+}}
\varsigma^{p/2-2} \exp(- \varsigma\| x - \theta
\|^{2}/2)G(d\varsigma)}{\int_{\R_{+}} \varsigma^{p/2}
\exp(- \varsigma\| x - \theta\|^{2}/2) G(d\varsigma)}\nonumber\\
&&\hspace*{139pt}\hphantom{\int_{\R^{p}}  \biggl(}{} + \gamma
^{2}(x)  \biggr)  \\
&&\hphantom{\int_{\R^{p}}}{} \cdot \int_{\R_{+}}  \biggl( \frac{\varsigma}{2 \pi}
\biggr)^{p/2}\nonumber\\
&&\hspace*{41.5pt}{}\cdot
\exp \biggl( - \frac{\varsigma}{2} \|x - \theta\|^{2}  \biggr)
G(d\varsigma) \, dx.
\nonumber
\end{eqnarray}}}%
Now, through a monotone likelihood ratio argument, the ratio of
integrals in (\ref{betatheta})
can be seen to be bounded from below by the constant $k$ in (\ref
{mixsuff2}). Hence the inequality in
(\ref{mixsuff2}) gives
\begin{eqnarray*}
\beta(\theta) &\leq&\int_{\R^{p}} \bigl(k \bigtriangleup\gamma(x) +
\gamma^{2}(x)\bigr)\\
&&\hphantom{\int_{\R^{p}}}{}\cdot \int_{\R_{+}}  \biggl( \frac{\varsigma}{2 \pi}
 \biggr)^{p/2}\\
&&\hphantom{\int_{\R^{p}}{}\cdot\int_{\R^{p}}}{}\cdot
\exp \biggl( - \frac{\varsigma}{2} \|x - \theta\|^{2}  \biggr)
G(d\varsigma) \, dx \\  &<&   0   .
\end{eqnarray*}

Finally, remembering that $\alpha(\theta)$ is nonpositive, it follows
that the risk difference
$\alpha(\theta) + \beta(\theta)$ between~$\delta_0$ and~$\delta
_{0}-\gamma$ is negative,
which proves the theorem.
\end{pf*}

The improved loss estimator result in Theorem~\ref{th3.1} for scale
mixture of normal
distributions family was extended to a more general family of spherically
symmetric distributions in \cite{FourdrinierWells1995b}.\vadjust{\goodbreak} In this
setting the conditions for improvement rest on the generating
function $g$ of the spherical density $p_{\theta}$. A sufficient
condition for domination of $\delta_0$ has the usual form $k \Delta
\gamma+ \gamma^{2} \leq0$.

\begin{theorem}\label{th3.2}
Assume the spherical distribution of $X$ with
generating function $g$ has finite fourth moment. Assume the
function $\gamma$ is nonnegative and twice weakly differentiable on
$\R^{p}$ and satisfies\break $E_{\theta}[\gamma^{2}] < \infty$. If, for
every $s \geq0$,
\begin{equation}
\frac{\int^{\infty}_{s} g(z) \, dz}{2   g(s)} \leq\frac{\delta_0}{p}
\end{equation}
and if there exists a constant $k$ such that, for any $s
\geq0$,
\begin{equation} 0 < k < \frac{\int^{\infty}_{s} zg(z)\,dz - s
\int^{\infty}_{s} g(z)\,dz}{2g(s)},
\end{equation}
then a
sufficient condition for $\delta_0 - \gamma$ to dominate $\delta_0$
is that $\gamma$ satisfies the differential inequality
\[
k   \Delta\gamma+ \gamma^{2} < 0   .
\]
\end{theorem}

We have shown that one can dominate the unbiased constant estimator
of loss by a shrinkage-type estimator. As in the normal case one may
wish to add a
frequentist-validity constraint to the loss estimation problem. It is
easy to show that
the only frequentist valid estimator of the form $\delta_{0}$ would
be the only frequentist valid loss estimator. The proof of this
result follows from a randomization of the origin technique as in
Hsieh and Hwang \cite{HsiehHwang1993}.

%




\subsection{Estimating the Quadratic Loss of the Mean of a Spherical
Distribution with a
Residual Vector} \label{subsection4.2}

In this  section, we extend the ideas of the previous sections
to a spherically symmetric distribution with a residual vector.
We first develop an unbiased estimator of the loss and then construct
a dominating shrinkage-type estimator. An
important feature of our results is that the proposed loss estimates
dominate the unbiased estimates for the entire class of spherically
symmetric distributions. That is, the domination results are robust
with respect to spherical symmetry.

Let $(X,U) \sim \operatorname{SS}(\theta,0)$ where $\operatorname{dim}   X =
\operatorname{dim}   \theta= p$ and
$\operatorname{dim}   U = \operatorname{dim}   0 = k$ ($p + k =
n$). For convenience of notation, here
$(X,U)$ and $(\theta,0)$ represent $n \times1$ vectors (see Appendix
\ref{A.2} for more
details on this mo\-del). Unlike  Section \ref{subsection4.1},
the dimension of the observable $(X,U)$ is greater than the dimension
of the estimand~$\theta$.
This model arises as the canonical form of the following seemingly more
general model, the
general linear model. Let $V$ be an $n \times p$ matrix (of full rank
$p$) which is often referred
to as the design matrix. Suppose an $n \times1$ vector $Y$ is observed
such that
$Y = V \beta+ \varepsilon$ where $\beta$ is a $p \times1$ vector of
(unknown) regression
coefficients and $\varepsilon$ is an $n \times1$ vector with a
spherically symmetric
distribution about $0$. A common alternative representation of this
model is
$Y = \eta+ \varepsilon$ where $\varepsilon$ is as above and~$\eta$
is in the column space
of $V.$

To understand this representation in terms of the general linear model, let
$G = (G_1^t,G_2^t)^t$
be an $n \times n$ orthogonal matrix partitioned such that the first
$p$ rows of $G$
(i.e., the rows of $G_1$ considered as column vectors) span the column
space of $V$. Now let
\[
\pmatrix{
X \cr
U
}
=
\pmatrix{
G_1 \cr
G_2
}Y =
\pmatrix{
G_1 \cr
G_2
}
V \beta+ G \varepsilon=
\pmatrix{
\theta\cr
0
}
+ G \varepsilon
\]
with $\theta= G_1 V \beta$ and $G_2 V \beta= 0$ since the rows of
$G_2$ are orthogonal
to the columns of $V$. It follows from the definition that $(X,U)$
has a spherically symmetric distribution about $(\theta,0)$.
In this sense, the model given above is the canonical form of
the general linear model.

The usual estimator of $\theta$ is the orthogonal projector
$X$. A class of competing point estimators which are
also considered is of the form $\varphi= X -
\| U \|^{2}g(X)$;
$g$ is a measurable function from $\R^p$ into $\R^p$. This
class of estimators is closely related to Stein-like estimators
(when estimating the mean of a normal
distribution, the square of the residual term
$\| u\|$ is used as an estimate of the unknown variance). Their
domination properties are robust with respect to spherical symmetry
(cf. \cite{CellierFourdrinier1995} and \cite{CellierFourdrinierRobert1989}).
We will first consider estimation
of the loss of the usual least squares estimator $X$, then
estimation of the loss of the more general shrinkage estimator
$\varphi$. In order to assure the finiteness of their risk of the
usual estimator $X$ and the risk of the shrinkage
estimator $\varphi$, we need two hypotheses (H1) and (H2) given in
\cite{CellierFourdrinier1995}.

In the spherical case in Section \ref{Section3}, the risk of $X$ was constant
with respect to $\theta$. Thus this risk provides an
unbiased estimator of the loss, that is, $p E[R^{2}]/n$,
which is subject to the knowledge of
$E[R^{2}]$. Its properties, as the properties of any improved
estimator, may depend on the specific underlying distribution. An
important feature of the results in this subsection is that we
propose an unbiased estimator $\delta_{0}$ of the loss of
$X$ which is available for every spherically symmetric
distribution (with finite fourth moment), that is,
$\delta_{0}(X,U) = p   \|U \|^{2} / k$.
Thus we do not need to know
the specific distribution, and we get robustness with an estimator
which is no longer constant. Notice $\delta_{0}$ makes sense
because $p < n$ (i.e., $k \geq1$).

In this subsection, we consider estimation of $\theta$ by $X$ so that,
as in the work of Fourdrinier and Wells \cite{FourdrinierWells1995},
we deal with estimating the loss
$\|X - \theta\|^2$. An unbiased estimator of that loss is given by
$\delta_{0}(X,U) = p   \|U\|^{2}/k$, that we write $\delta_{0}(U)$
since it
depends only on $U$. The unbiasedness of $\delta_{0}$ follows from Corollary
\ref{corollaryA} by taking $q = 0$ and $\gamma\equiv1$.
The goal of this subsection is to prove the domination of the
unbiased estimator $\delta_{0}$ by a competing estimator~$\delta$ of
the form
%
\begin{equation} \label{improvedlamda}
\delta(X,U) = \delta_{0}(U) - \|U \|^{4}   \gamma(X),
\end{equation}
where $\gamma$ is a nonnegative function. It is important to
notice that the ``residual term'' $\| U\|$ appears explicitly in
the shrinkage function. It has
been noted in~\cite{CellierFourdrinier1995} that the use of this
term allows fewer assumptions about the distributions than when it
does not appear. Specifically, this including gives a robustness
property to the results, since they are valid for the entire class
of spherically symmetric distributions.

We require the real-valued function $\gamma$ to be twice weakly differentiable,
in order to include basic examples, which are not twice differentiable.
The following domination result is given in \cite{FourdrinierWells1995}.
We will see below that it appears as a consequence of a more general result
when shrinkage estimators of $\theta$ are involved.

\begin{theorem} \label{theorem7.2}
Assume that $p \geq5$, the distribution of $(X,U)$
has a finite fourth moment and the function $\gamma$ is twice weakly
differentiable on $\R^p$ and there exists a constant $\beta$ such
that $\gamma(t)\leq\beta/\|t\|^2$. A sufficient condition under
which the estimator
$\delta$ in~(\ref{improvedlamda}) dominates the unbiased estimator
$\delta_{0}$ is that~$\gamma$ satisfies the differential inequality
%
\begin{equation} \label{improvedlamdadom}
\gamma^{2} + \frac{2}{(k + 4)(k + 6)} \bigtriangleup\gamma\leq0   .
\end{equation}
\end{theorem}

The standard example where $\gamma(t) = d/\|t \|^{2}$ for all
$t \neq0$ with $d > 0$ satisfies the conditions of the theorem.
More precisely, it is easy to deduce that
$\bigtriangleup\gamma(t) = - 2 \, d   (p - 4)/\|t \|^{4}$
and thus the sufficient condition of the
theorem is written as $0 < d \leq4   (p - 4)/(k + 4)(k + 6)$,
which only occurs when $p \geq5$. Straightforward calculus shows that
the optimal va\-lue of $d$ is given by $2   (p - 4)/(k + 4)(k +6)$.
The optimal constant in \cite{CellierFourdrinier1995} is equal to $2
  (p - 4)$.
The extra terms in the denominator compensate for the~$\|U\|^{4}$ term
in our estimator.

We now consider the estimation of the loss of a~class of
shrinkage estimators considered in \cite{CellierFourdrinier1995}
(with a slight modification of their form in order to have notations
coherent with those of the previous sections), that is,
location estimators of the form
%
\begin{eqnarray} \label{phig}
\varphi_{g} = X + \|U\|^{2}   g(X),
\end{eqnarray}
where $g$ is a weakly differentiable function from $\R^p$ into $\R^p$.
In \cite{CellierFourdrinier1995} it is shown that, if
$\|g\|^{2} \leq-2   \operatorname{div} g/\break(k + 2)$, then $\varphi_g$
dominates $X$,
under quadratic loss for all spherically symmetric distributions
with a finite second moment. A general example of a member of this
class of estimators is with
$ g(X) = \break- r(\|X\|^{2}) \frac{A(X)}{b(X)}$,
where $r$ is a positive differentiable and nondecreasing function,
$A$ is a positive definite symmetric matrix and $b$ is a positive
definite quadratic form of~$\R^p$. When $r$ is equal to some
constant $a$, $A$ is the identity on $\R^p$ and the quadratic form
$b$ is the usual norm, $g$ reduces to $a/ \| X \|^{2}$. It can be shown
that the optimal choice of $a$ equals $(p - 2)/(k +2)$. A~member of the
class is
$ \varphi_{r} = X - (p -2) \frac{\|U \|^{2}}{k + 2}\frac{X}{\|X\|^{2}}$,
the James--Stein estimator used when the variance is unknown as in
Section \ref{Section3}.

In Proposition 2.3.1 of Section 2.3 of \cite{CellierFourdrinier1995},
it is shown that an unbiased estimator of the loss of the shrinkage
estimator $\varphi_g$ is given by
%
\begin{eqnarray}\label{deltag0}
&& \delta^{g}_{0}(X,U) =\frac{p}{k} \|U\|^{2} + \biggl(\| g(X)\|^{2}\nonumber\hspace*{-20pt}
\\[-8pt]
\\[-8pt]
&&\hspace*{110pt}{} + \frac{2}{k + 2}   \operatorname{div}g(X) \biggr) \|U\|^{4}.
\nonumber\hspace*{-20pt}
\end{eqnarray}

As in Theorem \ref{theorem7.2} above, the unbiased estimator of the
loss can be
improved by a shrinkage estimator of the loss. Thus the competing
estimator we consider is
%
\begin{equation} \label{improvedlamdag}
\delta_\gamma^{g}(X,U) = \delta^{g}_{0}(X,U) - \| U\|^{4}   \gamma(X),
\end{equation}
where $\gamma$ is a nonnegative function. Note that
(\ref{improvedlamdag}) is a true shrinkage estimator, while
Johnstone's \cite{Johnstone1988} optimal loss estimate for the
normal case is an expanding estimator. This is not contradictory
since we are using a different estimator than Johnstone and he was
only dealing with the normal case. If $g \equiv0$, the following
result reduces to Theorem \ref{theorem7.2}.
%
\begin{theorem} \label{theorem7.3}
Assume that $p \geq5$, the distribution of $(X,U)$ has a finite
fourth moment and the function $\gamma$ is twice weakly
differentiable on $\R^p$ and there exists a constant $\beta$
such that $\gamma(t)\leq\beta/\|t\|^2$. A sufficient condition
under which the estimator $\delta_\gamma^{g}$ given in
(\ref{improvedlamdag}) dominates the unbiased estimator
$\delta^{g}_{0}$ is that $\gamma$ satisfies the differential
inequality
%
\begin{eqnarray} \label{4.15}
&&\gamma^{2} - \frac{4}{k + 2}   \gamma  \operatorname{div}g +
\frac{4}{k + 6}
\operatorname{div}(\gamma  g)\nonumber
\\[-8pt]
\\[-8pt]&& \quad {} + \frac{2}{(k + 4)(k + 6)}
\bigtriangleup\gamma\leq0   .
\nonumber
\end{eqnarray}
\end{theorem}

\begin{pf} Since the distribution of $(X,U)$ is spherically symmetric
around $\theta$, it suffices to obtain the result working
conditionally on the radius.
For $R > 0$ fixed, we can compute using the uniform distribution $U_{R,
\theta}$ on
the sphere $S_{R, \theta}$. Thus the conditional risk difference between
$\delta_\gamma^{g}$ and $\delta_0^{g}$, according to (\ref
{improvedlamdag}), equals
\begin{eqnarray*}
&& E_{R, \theta}  [(\delta_\gamma^{g}(X,U) -\|\varphi(X,U) -
\theta\|^{2})^{2} ]
\\[1pt]
&& \qquad {}-
E_{R, \theta}  [(\delta^{g}_{0}(X,U) - \| \varphi(X,U) -\theta
\|^{2})^{2}  ]
\\[1pt]
&&   \quad
= E_{R, \theta}  [ \|U\|^{8}   \gamma^{2}(X)  ]\\[1pt]
&&  \qquad {}
- E_{R, \theta} [ 2   \| U \|^{4} \gamma(X)\\[1pt]
&&  \hphantom{{}-E_{R, \theta} [}\qquad {}\cdot
(\delta^{g}_{0}(X,U) - \| \varphi(X,U) - \theta\|^{2})  ],
\end{eqnarray*}
that is, expanding and separating the integrand\break terms depending on
$\theta$,
\begin{eqnarray*}
&&E_{R, \theta}  \biggl[ \|U\|^{8}   \gamma^{2}(X)
- 2   \frac{p}{k}   \|U\|^6  \gamma(X)
\\[1pt]
&&\hspace*{54pt}{}- \frac{4}{k+2}   \|U\|^{8}   \operatorname{div}g(X)  \biggr]
 \\[1pt]
&&  \quad {}
+ E_{R, \theta}  [4   \|U\|^6   (X - \theta)^t \gamma(X)
g(X)  ]\\[1pt]
&& \quad {}
+ E_{R, \theta}  [2   \|U\|^4   \|X - \theta\|^2 \gamma(X)
 ] ,
\end{eqnarray*}
according to (\ref{deltag0}) (note that the two terms involving
$\|g(X)\|^2$ cancel).
Now we have
\begin{eqnarray*}
&&E_{R, \theta}  [4  \| U \|^{6}   (X -\theta)^t \gamma(X)
g(X)  ]\\[1pt]
&& \quad
= \frac{4}{k + 6} E_{R, \theta}  [\| U \|^{8}
  \operatorname{div}   ( \gamma(X)   g(X))  ]
\end{eqnarray*}
according to Lemma \ref{LemmaA5} and
\begin{eqnarray*}
&&E_{R, \theta}      [2   \|U\|^4   \|X - \theta\|^2 \gamma
(X)  ]
  \\[1pt]
  && \quad   =
E_{R, \theta}      \biggl[\frac{2   p}{k + 4} \| U \|^6   \gamma
(X) \\[1pt]
&&\hphantom{\quad   =
E_{R, \theta}      \biggl[}{}+
\frac{2}{(k + 4)(k + 6)}   \| U \|^{8}   \triangle\gamma(X)  \biggr]
\end{eqnarray*}
according to Corollary \ref{corollaryA}. Therefore the above
conditional risk difference
is equal to
\begin{eqnarray*}
&&E_{R, \theta}    \biggl[  \|U\|^{8}
   \biggl(
\gamma^{2}(X) - \frac{4}{k+2}   \operatorname{div}g(X) \\
&&\hphantom{E_{R, \theta}    \biggl[  \|U\|^{8}
   \biggl(}{}+
\frac{4}{k + 6}   \operatorname{div}   ( \gamma(X)   g(X))\\
&&\hphantom{E_{R, \theta}    \biggl[  \|U\|^{8}
   \biggl(}{} +
\frac{2}{(k + 4)(k + 6)}   \triangle\gamma(X)
   \biggr)    \biggr]
 \\
&&  \quad {}
+ E_{R, \theta}    \biggl[
2   p  \biggl(\frac{1}{k-4} - \frac{1}{k} \biggr) \|U\|^6  \gamma(X)
   \biggr]
\end{eqnarray*}
which is bounded above by the first expectation since the function
$\gamma$ is nonnegative.
Hence, the sufficient condition for domination is (\ref{4.15}) in
order that the inequality
$R(\delta^{g}, \theta,\varphi) \leq R(\delta^{g}_{0}, \theta,
\varphi)$
holds.
\end{pf}


\section{Discussion}
\label{discussion}

There are several areas of the theory of loss
estimation that we have not discussed. Our primary focus has been
on location parameters for the multivariate normal and spherical
distributions. Loss estimation for exponential families is addressed
in Lele \cite{Lele1992,Lele1993} and Rukhin
\cite{Rukhin1988}. In \cite{Lele1992} and \cite{Lele1993} Lele
developed improved loss estimators for point estimators in the
general setup of Hudson's \cite{Hudson1978} subclass of continuous
exponential family. Hudson's family essentially includes distributions for
which the Stein-like identities hold; explicit calculations and loss
estimators are given for the gamma distribution, as well as for improved
scaled quadratic loss estimators in the
Poisson setting for the Clevenson--Zidek \cite{ClevensonZidek1975}
estimator. Rukhin \cite{Rukhin1988} studied the posterior
loss estimator for a Bayes estimate (under quadratic loss) for the
canonical parameter of a linear exponential family.

As pointed out in the \hyperref[Section1]{Introduction}, in the known variance normal
setting, Johnstone \cite{Johnstone1988} used a version of Blyth's
lemma to show that the constant loss estimate $p$ is admissible if
$p\leq4$. Lele \cite{Lele1993} gave some additional sufficient
conditions for
admissibility in the general exponential family and worked out the
precise details for the Poisson model. Rukhin \cite{Rukhin1988}
considered loss functions for the simultaneous estimate of $\theta$
and $L(\theta,\varphi(X))$ and deduced some interesting admissibility results.

A number of researchers have investigated impro\-ved
estimators of a covariance matrix, $\Sigma$, under the Stein loss,
$L_{S}(\hat{\Sigma}, \Sigma) =
\operatorname{tr}(\hat{\Sigma} \Sigma^{-1}) - \log|\hat{\Sigma} \Sigma^{-1}| -
p $, using an unbiased
estimation of risk technique. In the normal case,
\cite{DS85,Haff79JMA,S77a,S77b}, and \cite{T84} proposed improved estimators that
dominate the sample covariance under~$L_{S}(\hat{\Sigma}, \Sigma)$. In
\cite{KubokawaSrivastava99}, it was shown that the domination of these
improved estimators
over the sample covariance estimator is robust with respect to the
family of elliptical
distributions. To date, there has not been any work on improving the unbiased
estimate of $L_{S}(\hat{\Sigma}, \Sigma)$.

In addition to the theoretical ideas discussed in the previous
sections there are very practical applications of loss estimation.
The primary application of loss estimation ideas is to model
selection. It was shown by Fourdrinier and Wells
\cite{FourdrinierWells1994} that improved loss estimators give more
accurate model selection procedures. Bartlett, Boucheron and
Lugosi~\cite{Bartlettetal2002} studied model selection strategies based on
penalized empirical loss minimization and pointed out the equivalence
between loss estimation and da\-ta-based complexity penalization. It
was shown that any good loss estimate may be converted into a
data-based penalty function and the performance of the estimate is
governed by the quality of the loss estimate. Furthermore, a
selected model that minimizes the penalized empirical loss achieves
an almost optimal trade-off between the approximation error and the
expected complexity, provided that the loss estimate on which the
complexity is based is an approximate upper bound on the true loss.
The key point to stress is that there is a fundamental dependence on
the notions of good complexity regularization and good loss
estimation. The ideas in this review lay the theoretical foundation
for the construction of such loss estimators and model selection rules
as well as give a decision-theoretic analysis of their statistical
properties.

In linear models the notion of degrees of \mbox{freedom} plays the
important role as a model complexity measure in various model
selection criteria, such as Akai\-ke information criterion (AIC)
\cite{Akaike1973} , Mallow's~$C_p$~\cite{Mallows1973}, and Bayesian
information criterion (BIC) \cite{Schwartz1978}, and generalized
cross-validation (GCV) \cite{CravenWahba1979}. In regression the
degrees of freedom are the trace of the so-called ``hat'' matrix.
Efron \cite{Efron2004} pointed out that the theory of Stein's
unbiased risk estimation is central to the ideas underlying the
calculation of the degrees of freedom of certain regression
estimators.

Specifically, let $Y$ be a random vector having an~$n$-variate normal
distribution
${\cal N}   (\theta,  \sigma^{2}   I_n)$
with unknown $p$-dimensional mean $\theta$ and identity covariance
matrix $\sigma^{2} I_n$.
Let $\hat{\theta}=\varphi(Y)$ be an estimate of $\theta$.
In regression one focuses on how accurate $\varphi$ can be in
predicting using a new response vector $y^{\mathrm{new}}$. Under the
quadratic loss, the prediction risk is $E\{\|Y^{\mathrm{new}} - \theta
\|^{2}\}/n$. Efron \cite{Efron2004} noted that
%
\begin{eqnarray} \label{efronrisk}
E\{\| \varphi- \theta\|^{2}\} &=& E \{ \| Y - \varphi(Y) \|^{2} - n
\sigma^{2}\} \nonumber
\\[-8pt]
\\[-8pt]&&{}+
2 \sum^{n}_{i=1}   \operatorname{Cov} (\varphi_{i}, Y_{i}).
\nonumber
\end{eqnarray}
This expression suggests a natural definition of the degrees of freedom
for an estimator $\varphi$ as $\operatorname{df}(\varphi) = \sum^{n}_{i=1}
\operatorname{Cov} (\varphi_{i}, Y_{i})/\sigma^{2} =
E_\theta[(Y-\theta)^t\varphi(Y)]/\sigma^{2}.$ Thus one can define a
$C_{p}$-type quantity
%
\begin{eqnarray} \label{cp}
C_{p}(\varphi) = \frac{\| Y - \varphi\|^{2}}{n} +
\frac{2\operatorname{df}(\varphi)}{n} \sigma^{2}
\end{eqnarray}
which has the same expectations as the true prediction error but may
not be an
estimate if $\operatorname{df}(\varphi)$ and $\sigma^{2}$ are unknown. However, if
$\varphi$ is weakly
differentiable and $\hat{\sigma}^{2}$ is an unbiased estimate of
$\sigma^{2}$,
the integration by parts formula in Lemma \ref{LemmaA1}
implies that $\operatorname{df}(\varphi)  \sigma^{2}= E_\theta[\operatorname
{div}\varphi(Y)   \hat{\sigma}^{2}]$, hence $\operatorname
{div}\varphi  \hat{\sigma}^{2}$ is unbiased
estimate for the complexity parameter term, $\operatorname{df}(\varphi)   \sigma
^{2}$, in (\ref{cp}).
Therefore an unbiased estimate for the prediction error is
%
\begin{eqnarray} \label{cp*}
C^{*}_{p}(\varphi) = \frac{\|Y- \varphi\|^{2}}{n} + \frac{2
\operatorname{div}\varphi}{n} \hat{\sigma}^{2}.
\end{eqnarray}

Note that, if $\varphi$ is a linear estimator ($\varphi= \mathbf{S}
y$ for some matrix $\mathbf{S}$
independent of $Y$), then it is easy to show that this definition
coincides with the definition
of generalized degrees of freedom given by Hastie and Tibshirani
\cite{HastieTibshirani1990} since $\operatorname{div}\varphi= \operatorname{tr}(\mathbf{S})$. Note that, if $\varphi$
also depends on $\hat{\sigma}^{2}$, then (\ref{efronrisk}) needs to
be augmented by additional
derivative terms with respect to $\hat{\sigma}^{2}$ as in Theorem
\ref{unknownvar}.

Other approaches for estimating the complexity term penalty involve
the use of resampling methods \cite{Efron2004,Ye1998} to
directly estimate the prediction error. A $K$-fold cross-validation randomly
divides the original sample into $K$ parts, and rotates through each part
as a test sample and uses the remainder as a training sample.
Cross-validation provides an approximately unbiased estimate of the prediction
error, although its variance can be large. Other commonly used
resampling techniques are the nonparametric and parametric bootstrap methods.

A number of new regularized regression methods have recently been
developed, starting with
Ridge regression
\cite{HoerlKennard1970}, followed by the Lasso
\cite{Tibshirani1996}, the Elastic Net \cite{ZouHastie2005}, and
LARS \cite{LARS2004}. Each of these estimates is weakly
differentiable and has the form of a general shrinkage estimate;
thus the prediction error estimate in (\ref{cp*}) may be applied to construct
a model selection procedure. Zou, Hastie and
Tibshirani \cite{ZouHastieTibshirani2007} used this idea to develop a
model selection method for the Lasso. In some situations verifying the
weak differentiability of $\varphi$ may be complicated.

Loss estimates have been used to derive nonparametric penalized
empirical loss estimates in the context of function estimation,
which adapt to the unknown smoothness of the function of interest.
See Barron et al. \cite{BarronBirgeMassart1999} and Donoho and
Johnstone \cite{DonohoJohnstone1995} for more details.

In the previous sections, the usual quadratic loss
$L(\theta,\varphi(x)) = \|\varphi(x)-\theta\|^2$ was considered to
evaluate various
estimators $\varphi(X)$ of $\theta$. The squared norm $\|x-\theta
\|^2$ was crucial in the
derivation of the properties of the loss estimators in conjunction with
its role in the
normal density or, more generally, in a~spherical density. Other losses
are thinkable but, to
deal with tractable calculations, it matters to keep the Euclidean norm
as a component of the
loss in use. Hence a natural extension is to consider losses which are
a function of
$\|x-\theta\|^2$, that is, of the form $c(\|x-\theta\|^2)$ for a
nonnegative function $c$
defined on $\R_{+}$. The problem of estimating a~function $c$ of
$\|x-\theta\|^2$ was tackled by
Fourdrinier and Lepelletier~\cite{FourdrinierLepelletier2008} to which
we refer the reader for more details. In particular, they focused on
the fact
that estimating $c(\|x-\theta\|^2)$ can be viewed as an evaluation
of a quantity which is not necessarily a loss. Indeed it includes the
problem of
estimating the confidence statement of the usual confidence set
$\{ \theta\in\R^{p}   \mid \|x-\theta\|^2 \leq c_{\alpha}\}$
with confidence coefficient $1-\alpha$: $c$ is the indicator function
$\one_{[0,c_{\alpha}]}$.

\begin{appendix}
\section*{Appendix}
\label{appendix}


\subsection{Risk Finiteness Conditions}\label{A.1}

\begin{lemma} \label{LemmaA2}
\textup{1.} Let $X \sim{\cal N}(\theta, I_p)$, where $\theta$ is unknown, and
denote by $E_\theta$
the expectation with respect to the distribution of $X$. Consider an
estimator of $\theta$ of the
form $\varphi(X) = X +   g(X)$ where $g$ is a~function from $\R^p$
into $\R^p$.

\textup{a.} If $g$ is such that $E_\theta[\|g(X)\|^2] < \infty$, then the
quadratic risk of
$\varphi(X)$, that is, $R(\theta,\varphi) = E_\theta[\|\varphi
(X)-\theta\|^2]$, is finite.

\textup{b.}
 If, in addition, the function $g$ is weakly differentiable so
that $ \delta_0(X) = p + 2   \operatorname{div}g(X) + \|g(X)\|^2 $
is an
unbiased estimator of the loss $\|\varphi(X)-\theta\|^2$, then the
risk of $\delta_0(X)$
defined by
${\cal R}(\theta,\varphi, \delta_0) = E_\theta[(\delta_0(X) -
\|\varphi(X)-\theta\|^2)^2]$
is finite as soon as $E_\theta[\|g(X)\|^4] < \infty$ and $E_\theta
[(\operatorname{div}g(X))^2] < \infty$.

 \textup{2.}  Let $X \sim{\cal N}(\theta,\sigma^2 I_p)$, where $\theta$ and
$\sigma^2$ are unknown, let
$S$ be a nonnegative random variable independent of $X$ and such that
$S \sim\sigma^2 \chi_n^2$ and denote by $E_{\theta,\sigma^2}$ the
expectation with respect to
the joint distribution of $(X,S)$. Consider an estimator of $\theta$
of the form
$\varphi(X,S) = X + S   g(X,S)$ where $g$ is a function from $\R^p
\times\R_+$ into $\R^p$.

\textup{a.} If $g$ is such that $E_{\theta,\sigma^2}[S^2   \|g(X,S)\|^2] <
\infty$, then the quadratic
risk of $\varphi(X)$, that is,
$R(\theta, \sigma^2, \varphi) = E_{\theta,\sigma^2}[\|\varphi
(X,S)-\theta\|^2 / \sigma^2]$, is
finite.

\textup{b.} If, in addition, the function $g$ is weakly differentiable so that
\begin{eqnarray*}
&&\delta_0(X,S)\\
 && \quad = p + S \biggl \{(n+2)   \|g(X,S)\|^2 \\
 &&\hphantom{\quad = p + S \biggl \{}{}+ 2
\operatorname{div}_Xg(X,S)
+ 2   S \frac{\partial}{\partial S} \|g(X,S)\|^2 \biggr\}
\end{eqnarray*}
is an unbiased estimator of the loss $\|\varphi(X,S) -\break \theta\|^2/
\sigma^2$, then the risk of
$\delta_0(X,S)$ defined by
${\cal R}(\theta, \sigma^2,\break  \varphi, \delta_0) =
E_{\theta,\sigma^2}[(\delta_0(X,S) - \|\varphi(X,S)-\theta\|^2 /
\sigma^2))^2]$
is finite as soon as $E_{\theta,\sigma^2}[S^2   \|g(X,S)\|^4] <
\infty$,\break 
$E_{\theta,\sigma^2} [(S   \operatorname{div}g(X,S))^2] < \infty$ and
$E_{\theta,\sigma^2}     [ (S^2   \frac{\partial
}{\partial S}
\|g(X,\break S)\| )^2 ]$.
\end{lemma}
\begin{pf}
\textup{1.a.} The loss of $\varphi(X)$ can be expanded as
%
\begin{eqnarray}\label{equaA1}
\|\varphi(X)-\theta\|^2 &=& \|X-\theta\|^2 + \|g(X)\|^2\nonumber
\\[-8pt]
\\[-8pt] &&{}+ 2 (X-\theta
)^t g(X)   .
\nonumber
\end{eqnarray}
Now we have $E_\theta[\|X-\theta\|^2] = p < \infty$. Hence, by\break
Schwarz's inequality, it follows
from (\ref{equaA1}) that\break 
$|E_\theta[(X-\theta)^t g(X)]| \leq
(E_\theta[\|X-\theta\|^2])^{1/2}\cdot \break  (E_\theta[\|g(X)\|^2])^{1/2}$.
Therefore, as soon as\break 
$E_\theta[\|g(X)\|^2] < \infty$, we will have $|E_\theta[\|\varphi
(X)-\theta\|^2] < \infty$.
This is the desired result.
{\smallskipamount=0pt
\begin{longlist}[2.a.]
\item[b.] Note that, under the usual domination condition, that is,
$2   \operatorname{div}g(x) + \|g(x)\|^2 \leq0$ for any $x \in R^p$,
of $\delta_0(X)$ over $X$, the condition
$E_\theta[(\operatorname{div}g(X))^2] < \infty$
implies that $E_\theta[\|g(X)\|^4] < \infty$. We will have
${\cal R}(\theta,\varphi,\break \delta_0) = E_\theta[(\delta_0(X) -
\|\varphi(X)-\theta\|^2))^2]
< \infty$ as soon as $E_\theta[\delta_0^2(X)] < \infty$ and
$E_\theta[\|\varphi(X)-\theta\|^4] < \infty$.
Now\break $E_\theta[\delta_0^2(X)] = E_\theta[(p + 2   \operatorname
{div}g(X) + \|g(X)\|)^2] < \infty$
sin\-ce $E_\theta[(\operatorname{div}g(X))^2] < \infty$ and \mbox{$E_\theta
[\|g(X)\|^4] < \infty$}. Also
according to (\ref{equaA1})
\begin{eqnarray*}
 E_\theta[\|\varphi(X)-\theta\|^4] &=&
E_\theta[(\|X-\theta\|^2 + \|g(X)\|^2\\
&&\hspace*{9pt}\hphantom{E_\theta[}{} + 2 (X-\theta)^t g(X))^2]\\
&<& \infty
\end{eqnarray*}
since $E_\theta[\|X-\theta\|^4] < \infty$ and $E_\theta[\|g(X)\|^4]
< \infty$ and,
consequently, since $|(X-\theta)^t g(X)| \leq\|X-\theta\|
\|g(X)\|$ implies that
\begin{eqnarray*}
&&E_\theta[|(X-\theta)^t g(X)|^2] \\
&& \quad \leq E_\theta[\|X-\theta\|^2
\|g(X)\|^2] \\
&& \quad \leq  (E_\theta[\|X-\theta\|^4] )^{1/2}  (E_\theta
[\|g(X)\|^4] )^{1/2}
\end{eqnarray*}
by Schwarz's inequality.

\item[2.a.] Parallel to the case where the variance $\sigma^2$ is known, it
should be noticed that the
corresponding domination condition of $\delta(X,S)$ over $\delta
_0(X,S)$, that is, for any
$x \in R^p$ and any $s \in\R_+$,
$(n+2)   \|g(x,s)\|^2 + 2   \operatorname{div}_xg(x,s)
+ 2   s \frac{\partial}{\partial s} \|g(x,s)\|^2 \leq0$,
entails that the two conditions $E_{\theta,\sigma^2}[(S
\operatorname{div}g(X,S))^2] < \infty$
and\break 
$E_{\theta,\sigma^2}     [ (S^2
\frac{\partial}{\partial S}\|g(X,S)\| )^2 ]$
imply the condition\break 
$E_{\theta,\sigma^2}[S^2   \|g(X,S)\|^4] <
\infty$. Also the derivation
of the finiteness of $R(\theta, \sigma^2, \varphi)$ follows a
similar way as in~1.a.

\item[b.] We will have
${\cal R}(\theta, \sigma^2, \varphi, \delta_0) =
E_{\theta,\sigma^2}[(\delta_0(X,\break S) - \|\varphi(X)-\theta\|^2 /
\sigma^2))^2] < \infty$
as soon as $E_{\theta,\sigma^2}[(\delta_0(X,\allowbreak S))^2 < \infty$ and
$E_{\theta,\sigma^2}[\|\varphi(X)-\theta\|^4] < \infty$.
Now\break 
$
E_{\theta,\sigma^2}[(\delta_0(X,S))^2 =
E_{\theta,\sigma^2}[p + S  \{(n+2)   \|g(X,S)\|^2 + 2
\operatorname{div}_Xg(X,S)
+ 2   S \frac{\partial}{\partial S} \|g(X,S)\|^2 \}]
< \infty
$
since we assume that $E_{\theta,\sigma^2}[(S   \operatorname
{div}_Xg(X,S))^2] < \infty$ and\break 
$E_{\theta,\sigma^2}[S^2   \|g(X,S)\|^4] < \infty$. Also
$E_{\theta,\sigma^2}[\|\varphi(X,S)-\break\theta\|^4] =
E_{\theta,\sigma^2}[(\|X-\theta\|^2 + S^2 \|g(X,S)\|^2 + 2 S
(X-\break\theta)^t g(X,S))^2])^2]
< \infty$
since $E_\theta[\|X-\theta\|^4] < \infty$ and
$E_{\theta,\sigma^2}[S^2   \|g(X,S)\|^4] < \infty$
(note that $|(X-\theta)^t g(X,\allowbreak S)| \leq\|X-\theta\|   \|g(X,S)\|$
implies that
\begin{eqnarray*}
&&E_{\theta,\sigma^2}[|(X-\theta)^t S   g(X,S)|^2]\\
 && \quad \leq
E_{\theta,\sigma^2}[\|X-\theta\|^2   S^2   \|g(X,S)\|^2] \\
&& \quad \leq  (E_{\theta,\sigma^2 }[\|X-\theta\|^4] )^{1/2}
 (E_{\theta,\sigma^2}[S^2   \|g(X,S)\|^4] )^{1/2}
\end{eqnarray*}
by Schwarz's inequality).\hfill\qed
\end{longlist}}
\noqed\end{pf}


\subsection{Additional Technical Lemmas} \label{A.2}

This Appendix gives some technical results used in  Section \ref
{subsection4.2}.
The first two results deal with expectations conditioned on the radius
of a
spherically symmetric distribution in $\R^p\times\R^k$ centered at
$(\theta,0)$ where $\theta\in\R^p$. These expectations reduce to integrals
with respect to the uniform distribution $U_{R,\theta}$ on the sphere
\begin{eqnarray*}
S_{R,\theta}&=& \{y=(x,u)\in\R^p\times\R^k|\\
&&\hphantom{\{} (\|x-\theta\|^2
+\|u\|^2 )^{1/2}=R \}   .
\end{eqnarray*}
If $E_{R,\theta}[\psi]$ is the expectation
of some function $\psi$ with respect to $U_{R,\theta}$, the
expectation with
respect to the entire distribution is given by\vadjust{\goodbreak} $E_\theta[\psi]
=E [E_{R,\theta}[\psi] ]$ where $E$ is the expectation
with respect
to the distribution of the radius.

When the spherical distribution has a density with respect to the Lebesgue
measure, it is necessarily of the form $f (\|x-\theta\|^2+\|u\|
^2 )$
for some function $f$. Then the radius has density $R\rightarrow\sigma_{p+k}
f(R^2)R^{p+k-1}$ where $\sigma_{p+k}=\fraca{2\pi^{p+k}}{\Gamma (
\frac{p+k}{2} )}$. Therefore the expectation of any function
$\psi$ can
be written as
\[
E_\theta[\psi] = \int^\infty_0 \biggl[\int_{S_{R,\theta}}\psi(y)
U_{R,\theta}(dy) \biggr]f(R)\,dR .
\]

Note that for a vector $y=(x,u)\in S_{R,\theta}$, we have $x=\pi(y)$ and
$\|u\|^2=R^2-\|\pi(y)-\theta\|^2$ where $\pi$ is the orthogonal
projector from
$\R^p\times\R^k$ onto $\R^p$. Under~$U_{R,\theta}$, the distribution
$\pi(U_{R,\theta})$ of this projector has a~density with respect to the
Lebesgue measure on~$\R^p$ given by
$x\rightarrow
C^{p,k}_R (R^2-\|x-\theta\|^2 )^{\fraca{k}{2}-1}
\mathbf{1}_{B_{R,\theta}}(x)$ where
$C^{p,k}_R=\fracc{\Gamma (\frac{p+k}{2} )R^{2-p-k}}
{\Gamma (\frac{k}{2} )\pi^{p/2}}$ and $\mathbf
{1}_{B_{R,\theta}}$ is the
indicator function of the ball
$B_{R,\theta}= \{x\in\R^p|\|x-\theta\|\le R \}$ of radius
$R$ centered at $\theta$ in $\R^p$.

According to the above, as a spherically symmetric distribution on $\R^p$
around $\theta$, the radius of $\pi(U_{R,\theta})$ has density
\begin{eqnarray*}
r&\rightarrow&\sigma_p C^{p,k}_R  (R^2-r^2 )^{\fraca{k}{2}-1}
\mathbf{1}_{]0,R[}(r) r^{p-1}\\
&=&\frac{2R^{2-p-k}}
{B (\fraca{p}{2},\fraca{k}{2} )}
r^{p-1} (R^2-r^2 )^{\fraca{k}{2}-1}\mathbf{1}_{]0,R[}(r).
\end{eqnarray*}
We use repeatedly the fact that any such
projection onto a space of dimension greater than 0 and less than $p+k$ is
spherically symmetric with a density. Then we also often make use of its
radial density.
%
\begin{lemma} \label{LemmaA5}
For every twice weakly differentiable function $g(\R^p \rightarrow
\R^p)$ and for every function $h(\R_{+} \rightarrow\R)$,
%
\begin{eqnarray}
&&E_{R, \theta}  [h(\|U\|^2)(X-\theta)^t g(X) ] \nonumber
\\[-8pt]
\\[-8pt]&& \quad =
E_{R, \theta}  \biggl[\frac{H(\|U\|^2)}{(\|U\|^2)^{\fraca{k}{2}-1}}
\operatorname{div} g(X) \biggr]   ,
\nonumber
\end{eqnarray}
where $H$ is the indefinite integral, vanishing at 0, of the function
$t \rightarrow \frac{1}{2}h(t) t^{\fraca{k}{2}-1}$.
\end{lemma}
\begin{pf}
We have
\begin{eqnarray*}
&&E_{R,\theta} [h(\|U\|^2)(X-\theta)^t g(X) ]\\
      && \quad =       C^{p,k}_R     \int_{B_{R,\theta}}
h(R^2 - \|x - \theta\|^2)(x - \theta)^t\\
&&\hphantom{\quad =       C^{p,k}_R     \int_{B_{R,\theta}}}{}\cdot g(x)
 (R^2 - \|x - \theta\|^2 )^{\fraca{k}{2}-1}\,dx\\
      && \quad =       C^{p,k}_R     \int_{B_{R,\theta}}
\bigl(\nabla H(R^2-\|x-\theta\|^2)\bigr)^t g(x)\,dx
\end{eqnarray*}
since
\begin{eqnarray*}
&&\hspace*{-2pt}\nabla H(R^2-\|x-\theta\|^2)\\
 &&\hspace*{-2pt} \quad = -2H'(R^2-\|x-\theta\|^2)(x-\theta)\\
&&\hspace*{-2pt} \quad = h(R^2-\|x-\theta\|^2)
 (R^2-\|x-\theta\|^2 )^{\fraca{k}{2}-1}(x-\theta) .
\end{eqnarray*}
Then, by the divergence formula,
\begin{eqnarray*}
&&E_{R,\theta} [h(\|U\|^2)(X-\theta)^t g(X) ]\\
 && \quad =
C^{p,k}_R
\int_{B_{R,\theta}} \operatorname{div}\bigl (H(R^2-\|x-\theta\|
^2)g(x) \bigr)\,dx \\
&& \qquad {}- C^{p,k}_R \int_{B_{R,\theta}} H(R^2-\|x-\theta\|^2)\operatorname
{div} g(x)\,dx.
\end{eqnarray*}
Now, if $\sigma_{R,\theta}$ denotes the area measure on the
sphe\-re~$S_{R,\theta}$, the divergence theorem insures that the first integral equals
\[
C^{p,k}_R \int_{S_{R,\theta}} \bigl(H(R^2-\|x-\theta\|^2)g(x)\bigr)^t
\frac{x-\theta}{\|x-\theta\|} \sigma_{R,\theta}(dx)
\]
and is null since, for $x\in S_{R,\theta}$, $R^2-\|x-\theta\|^2=0$ and
$H(0)=0$. Hence, in terms of expectation, we have
\begin{eqnarray*}
&&E_{R,\theta} [h(\|U\|^2)(X-\theta)^t g(X) ]\\
&& \quad = C^{p,k}_R \int_{B_{R,\theta}}
\frac{H(R^2-\|x-\theta\|^2)}{ (R^2-\|x-\theta\|^2 )^
{\fraca{k}{2}}-1} \operatorname{div} g(x)
 \\
 && \hphantom{C^{p,k}_R \int_{B_{R,\theta}}}\qquad {}\cdot(R^2-\|x-\theta\|^2 )^{\fraca{k}{2}-1}\,dx\\
 && \quad = E_{R,\theta} \biggl[\frac{H(\|U\|^2)}{(\|U\|^2)^
{\fraca{k}{2}-1}}\operatorname{div} g(X) \biggr]
\end{eqnarray*}
which is the desired result.
\end{pf}

\begin{corollary} \label{corollaryA}
For every twice weakly differentiable function
$\gamma(\R^p \rightarrow\R_{+})$ and for every integer~$q$,
\begin{eqnarray*}
&&E_{R, \theta}  [\| U \|^{q}   \| X - \theta\|^{2}
  \gamma(X)  ] \\
  && \quad  = \frac{p}{k + q}
E_{R, \theta}  [\| U \|^{q + 2}   \gamma(X)  ]\\
&& \qquad {}+ \frac{1}{(k + q)(k + q + 2)}
E_{R,\theta}  [\| U \|^{q + 4} \bigtriangleup\gamma(X)  ]
  .
\end{eqnarray*}
\end{corollary}
\begin{pf}
Take $h(t) = t^{q/2}$ and $g(x) = \gamma(x)   (x - \theta)$ and
apply Lemma \ref{LemmaA5}
twice.
\end{pf}
\end{appendix}

\section*{Acknowledgments}
We are grateful to Bill and Rob
Strawderman as well as Rajendran Narayanan for their helpful
suggestions and comments that greatly aided the revision of the
manuscript.
Dominique Fourdrinier gratefully acknowledges
the support of the ANR Grant 08-EMER-002.
Martin T. Wells gratefully\vadjust{\goodbreak} acknowledges the
support of NSF Grant 06-12031 and NIH Grant R01-GM083606-01.

%

\end{document}